\documentclass[aps,twocolumn,amsmath,amssymb,longbibliography,prb,10pt]{revtex4-2}
\usepackage[colorlinks=true,allcolors=blue,hypertexnames=false]{hyperref}
\usepackage{amsmath}
\usepackage{dsfont}
\usepackage{textcomp}
\usepackage{tabularx}
\usepackage{graphicx}
\usepackage{soul}
\usepackage{feynmp-auto}
\usepackage{footmisc}
\usepackage[export]{adjustbox}
\usepackage{wrapfig}
\usepackage{bbm}

\makeatletter
\def\@bibdataout@aps{%
 \immediate\write\@bibdataout{%
  @CONTROL{%
   apsrev42Control%
   \longbibliography@sw{%
    ,author="48",editor="1",pages="0",title="0",year="1"%
   }{%
    ,author="48",editor="1",pages="0",title="",year="1"%
   }%
  }%
 }%
 \if@filesw
  \immediate\write\@auxout{\string\citation{apsrev42Control}}%
 \fi
}%
\makeatother

\newcommand{\beg}{\begin{equation}}
\newcommand{\en}{\end{equation}}
\newcommand{\bp}{\mathbf p}
\newcommand{\bq}{\mathbf q}

\newcommand{\bk}{\mathbf k}
\newcommand{\br}{\mathbf r}

\newcommand{\bn}{\mathbf n}

\newcommand{\bE}{\mathbf E}

\newcommand{\ttaut}{\hat\tau_3}

\newcommand \bel  {\begin{align}}
\newcommand \enl  {\end{align}}

\newcommand{\eps}{\epsilon}
\newcommand{\veps}{\varepsilon}
\newcommand{\lam}{\lambda}

\newcommand{\up}{\uparrow}
\newcommand{\dn}{\downarrow}

\newcommand{\eref}[1]{Eq.~(\ref{#1})}

\renewcommand{\Im}{\mathrm{Im}}
\newcommand{\Tr}{\mathrm{Tr}\,}

\newcommand{\ttau}[1]{\hat\tau_{#1}}
\newcommand{\itt}{i\hat\tau_2}
\newcommand{\gh}{\hat g}
\newcommand{\gc}{\check g}

\newcommand{\hs}{\hat\sigma}
\newcommand{\dDh}{\delta\hat\Delta}

\renewcommand{\Im}{\mathrm{Im}}

\newcommand{\Gel}{\gamma_{\rm el}}
\newcommand{\tveps}{\tilde{\eps}}
\newcommand{\avg}[1]{\left\langle #1\right\rangle_\bn}
\newcommand{\Gcal}{{\cal G}}
\newcommand{\Fcal}{{\cal F}}

\newcommand{\Ycal}{{\cal Y}}

\def\XXint#1#2#3{{\setbox0=\hbox{$#1{#2#3}{\int}$}
     \vcenter{\hbox{$#2#3$}}\kern-.5\wd0}}

\begin{document}

\title{Quasiclassical theory of nonlinear response in $d$-wave superconductors}

\author{Aidan Steineman}
\affiliation{Department of Physics, Kent State University, Kent, Ohio 44242, USA}
\author{Samuel Awelewa}
\affiliation{Department of Physics, Kent State University, Kent, Ohio 44242, USA}
\author{Maxim Dzero}
\affiliation{Department of Physics, Kent State University, Kent, Ohio 44242, USA}
\date{\today}

\begin{abstract}
Motivated by recent THz pump-probe experiments in unconventional superconductors, in this paper we use Keldysh--Nambu quasiclassical theory of $d$-wave superconductors to analyze the third-harmonic generation.
A systematic perturbative solution of the out-of-equilibrium Eilenberger equation for the Keldysh propagator is developed with respect to the amplitude of an external electromagnetic field.
We show that in a clean superconductor and for a spatially uniform electromagnetic field third harmonic response is absent. 
In the presence of elastic impurity scattering which we treat using the self-consistent scattering matrix approximation in the Born limit we find that the third-harmonic current density is a sum of the Higgs (amplitude) mode and quasiparticle contributions. We compute their dependence on disorder, temperature, frequency and polarization of external radiation.
\end{abstract}
\maketitle

\section{Introduction}\label{sec:intro}
Optical probes of correlated electronic matter beyond linear response have recently become one of the major experimental tools for studying collective excitations \cite{ShimanoTsuji2020,Armitage2026}. The nonlinear electromagnetic response of superconductors allows one to probe their collective modes, associated with the amplitude (Schmid–Higgs) \cite{ASchmid1968,SooryakumarKlein1980,VarmaLit1,VarmaLit2,Measson2014,AndersonAllThat2015,ShimanoTsuji2020,Volovik2014,PekkerVarma2015} and phase (Carlson–Goldman) fluctuations of the order parameter \cite{CarlsonGoldman73,Volkov1975,Volkov1979,SchmidSchon1975,SchmidSchon1979,Kulik1981}. In superconductors described by the weak coupling (BCS) theory the Schmid-Higgs mode is not a true collective mode because its energy is degenerate with the energy of the two-particle excitation threshold. As a result, for small deviations from equilibrium it exhibits power-law Landau damping — its dynamics can still be probed on relatively short time scales \cite{VolkovKogan1973}. Only for sufficiently strong deviations from equilibrium the amplitude mode may exhibit undamped periodic oscillations \cite{Spivak2004,Enolski2005,Enolski2005a,Levitov2006,Yuzbashyan2006,Yuzbashyan2008,Yuzbashyan2015}.  In passing we note that the Schmid–Higgs mode becomes a true collective mode only in superconductors with strong electron-phonon coupling where the retardation effects shift the frequency of the Higgs mode below the edge of the two-particle continuum and as a result it becomes fully isolated \cite{DY2026}.

\begin{figure}
\includegraphics[width=\linewidth]{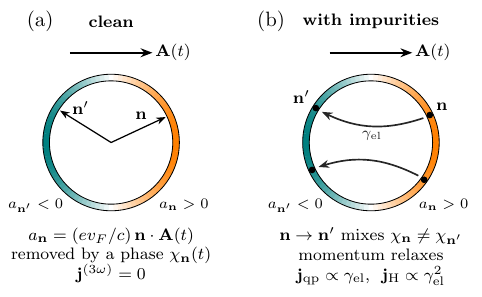}
\caption{Schematic representation of why the third harmonic generation requires momentum relaxation. The colors show the Doppler shift
$a_\bn(t)=(ev_F/c)\,\bn\cdot\mathbf A(t)$ of the quasiparticle energy on each trajectory $\bn$ of the Fermi surface
for a uniform vector potential $\mathbf A(t)$ along $x$, orange where the shift is positive and teal where it is
negative.
(a) Without impurities and for a linearized spectrum the field enters each trajectory only through its own phase
$\chi_\bn(t)$, with $\dot\chi_\bn=a_\bn$. The trajectories are coupled only through the pairing field $\Delta_\bn$, which is built
from equal-time functions in which these phases cancel. The electromagnetic field is therefore removed on every trajectory
(Appendix~\ref{AppendixA:gauge}), and the third-harmonic current vanishes. A momentum-dependent velocity (band
curvature) would also spoil this removal, but it lies outside the quasiclassical theory.
(b) Elastic impurity scattering $\bn\to\bn'$, with the single-particle rate $\Gel/2$, couples the trajectories through
the two-time self-energy, so the phase differences $\chi_{\bn'}-\chi_\bn$ remain and the field can no longer be
removed. Momentum relaxes, and for $\Gel\to0$ the quasiparticle and Higgs contributions scale as
$\mathbf j_{\rm qp}\propto\Gel$ and $\mathbf j_{\rm H}\propto\Gel^2$.}
\label{FigIntro}
\end{figure}

It is well established that the amplitude mode in weakly coupled superconductors is most directly accessed through third-harmonic generation, where it produces a resonant response when the drive frequency matches the order-parameter scale \cite{Silaev2019,MoorVolkovEfetov2017,Schwarz2020-Classification,Eremin2024}. 
As we have mentioned above, in conventional $s$-wave superconductors with the order parameter $\Delta$ in the weak coupling limit this sharp resonance appears at frequency $2\omega=2\Delta$  \cite{VolkovKogan1973,MoorVolkovEfetov2017,ChubukovHiggs2023,LiDzero2024,Nosov2025}, making the third-harmonic signal a clean probe of the amplitude mode. The amplitude mode oscillation $\delta{\Delta}_{2\omega}$ driven by external field is proportional to the product of the longitudinal pair susceptibility $\chi_{\textrm{SH}}(2\omega)$, which is singular at $2\omega=2\Delta$, and the square of external vector potential, ${\mathbf A}^2$ \cite{Tsuji2015,Cea2016}. These theoretical predictions have indeed been confirmed experimentally \cite{Shimano2014,Matsunaga2017,Shimano2013}. 

In addition to the third harmonic response from the Schmid-Higgs mode there is also a contribution from the charge density fluctuations, i.e. from the particle-hole channel, so that the third harmonic response of the current density in a superconductor can be represented as a sum of two distinct contributions
\beg\label{Decompose}
{\mathbf j}^{(3\omega)}(t)={\mathbf j}_{\textrm{SH}}^{(3\omega)}(t)+{\mathbf j}_{\textrm{CDF}}^{(3\omega)}(t),
\en
where ${\mathbf j}_{\textrm{SH}}^{(3\omega)}\propto \delta{\Delta}_{2\omega}$ accounts for the response due to the amplitude (Schmid-Higgs) mode fluctuations, while ${\mathbf j}_{\textrm{CDF}}^{(3\omega)}$ is governed mainly by charge density fluctuations in a superconductor. Whether the first term in \eqref{Decompose} dominates over the second one does not only depend on frequency of external radiation but also on the elastic disorder scattering rate \cite{Silaev2019,Aidan2026}. Disorder provides a key mechanism to relax the momentum thus allowing for coupling between the external radiation and quasiparticles, as we illustrate in Fig. \ref{FigIntro}. 

In the case of the $d$-wave superconductors, by contrast, the nodes in the order parameter remove this singularity \cite{BarlasVarma2013,Awelewa2026,Kazi2026} causing the order parameter amplitude oscillation to decay into nodal quasiparticles at every frequency. This has direct implication for the frequency dependence of the longitudinal pair susceptibility which is in the immediate vicinity of the Schmid-Higgs resonance is given by 
$\chi_{\textrm{SH}}(\nu)-\chi_{\textrm{SH}}(0)\sim\nu\log \nu$,  
where dimensionless parameter $\nu=|2\omega-2\sqrt{2}\Delta|/2\sqrt{2}\Delta$ has been assumed to be small \cite{Yuan2024,Kazi2026,VladIFE2026}. The extra factor of $\sqrt{2}$ appears here due to the normalized $d$-wave spectral factor $\sqrt{2}(n_x^2-n_y^2)$ ($\bn=\bp_F/p_F$ and $\bp_F$ is the Fermi momentum). 

Schwarz and Manske \cite{Manske-dwave2020} have studied the third-harmonic response in unconventional superconductors within the Anderson pseudospin formulation of BCS theory, with particular focus on $d$- and $d+s$-wave order parameters. 
In particular the current density was computed from 
\beg\label{ManskeCurrent}
{\mathbf j}(t)=e\sum\limits_\bk {\mathbf v}_{\bk-\frac{e}{c}{\mathbf A}(t)}n_{\mathbf k}(t), 
\en
where ${\mathbf A}(t)$ is an external vector potential, $\bk$ is the single particle momentum, ${\mathbf v}_\bk$ is the band velocity and $n_\bk(t)$ is the charge density, with the full lattice dispersion retained \cite{Manske-dwave2020}.
In their clean-limit analysis the coupling to the vector potential enters exclusively through the nonparabolicity of the band dispersion, and the charge-density-fluctuation contribution was found to exceed the Higgs contribution by more than two orders of magnitude. They also showed that the Higgs and phase-mode contributions are independent of the polarization of the driving field while the charge-fluctuation contribution is not. For a drive polarized along the nodal direction the transverse charge-fluctuation current vanishes and the divergent part of its longitudinal component cancels against the phase-mode contribution, so that only the Higgs term survives at that particular angle.

In this paper we use nonequilibrium field-theoretical technique to compute the third harmonic response in disordered $d$-wave superconductor. Technically, we solve Eilenberger equation for a quasiclassical propagator defined in Keldysh and Nambu spaces. The quasiclassical approach adopted here is complementary to the one developed in Ref. \cite{Manske-dwave2020}. In order to derive the quasiclassical equation one necessarily need to linearize the single-particle spectrum about the Fermi surface and integrate over the energy $\xi_{\mathbf p}$ with the density of states fixed at its Fermi-surface value. Our theory is therefore exactly particle--hole symmetric for any normal-state dispersion, and the band-curvature contributions that drive the response in Ref. \cite{Manske-dwave2020} are absent from it by construction. 
In exchange, it treats the nonequilibrium single-particle distribution function on the same footing as the condensate degrees of freedom. This is of particular importance for nodal superconductors: while in the $s$-wave case the low-energy quasiparticle processes are exponentially suppressed at low temperatures, which may justify working with an equilibrium single-particle distribution \cite{Tsuji2015,Cea2016,Seibold2021-Disorder,Wang2026}, in $d$-wave superconductors they are suppressed only as a power law, $\sim (T/\Delta)^2$, and the nonequilibrium distribution must be retained explicitly. The two approaches thus address complementary parts of the nonlinear response, and we compare them where they overlap, namely in the polarization dependence of the two channels.
The effects associated with particle-hole asymmetry can be included by perturbation theory \cite{VladIFE2026}.

The remainder of out paper is organized as follows. In the next Section we provide a brief summary of the quasiclassical formalism which we then use to determine the correction to the pairing amplitude which is proportional to the second power of the external vector potential. Section II is devoted to the discussion of the second harmonic response and longitudinal pair susceptibility. In Section IV we use our results from Section III to compute the third harmonic response of the current density. Sections V and VI are devoted to the discussion of our results and conclusions. Lastly, main technical details of the calculations are given in Appendices \ref{AppendixA:gauge}--\ref{app:third}.


\section{Model and formalism}\label{sec:model}
\subsection{Main equations}
\begin{figure}
\includegraphics[width=\linewidth]{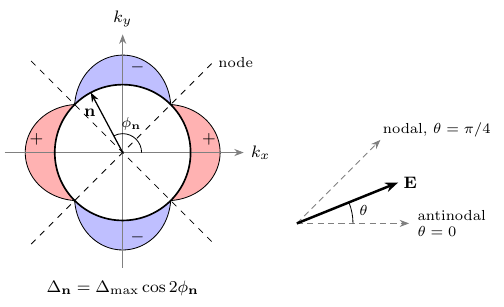}
\caption{Cylindrical Fermi surface (circle) and the $d$-wave order parameter
$\Delta_\bn=\Ycal(\phi_\bn)\Delta=\Delta_{\max}\cos2\phi_\bn$, $\Delta_{\max}=\sqrt2\,\Delta$.
The shaded lobes show $|\Delta_\bn|$, with $\Delta_\bn>0$ (red) along $k_x$ and $\Delta_\bn<0$ (blue) along $k_y$.
The dashed lines mark the nodes, $\phi_\bn=\pm\pi/4$ and $\pm3\pi/4$.
The angle $\phi_\bn$ of the direction $\bn$ on the Fermi surface and the angle $\theta$ of the linearly polarized
field $\bE$ are measured from the antinodal $x$ axis. The polarizations $\theta=0$ (antinodal) and
$\theta=\pi/4$ (nodal) are those of Fig.~\ref{Fig4}.}
\label{Fig1}
\end{figure}

We consider single band superconductor with cylindrical Fermi surface: 
\beg\label{Eq1}
\begin{aligned}
\hat{\cal H}&=\sum_{\bk,\sigma} \xi\left({\bk-\frac{e}{c}{\mathbf A}}\right) \hat{c}^\dagger_{\bk,\sigma}\hat{c}_{\bk,\sigma}\\&+
    \sum_{\bk,\bp,\bq} V_{\bk\bp} \hat{c}^\dagger_{\bk+\frac{\bq}{2},\uparrow} \hat{c}^\dagger_{-\bk+\frac{\bq}{2} ,\downarrow}\hat{c}_{-\bp+\frac{\bq}{2} ,\downarrow}\hat{c}_{\bp+\frac{\bq}{2},\uparrow},
\end{aligned}
\en
 where 
 $\hat{c}_{\bk\sigma}^\dagger$ ($\hat{c}_{\bk\sigma}$) are the creation (annihilation) fermionic operators, $\bk$ is momentum, $\sigma=\up,\dn$ is a spin,  
 $V_{\bk\bp}^{({\mathrm{d}})} $ is the pairing interaction,  $\xi(\bk)= k^2/2m - \veps_F$ is the single particle dispersion, $\veps_F$ is the Fermi energy, and $e$ is the electron charge; we set $\hbar=k_B=1$.  We project the interaction kernel $V_{\bk\bp}$ into the $d-$wave channel, $V_{\bk\bp}\to V_{\bk\bp}^{({\mathrm{d}})}$ which assumes a separable form  $V_{\bk\bp}^{(\mathrm{d})}=- (\lambda/\nu_F){\cal Y}(\phi_\bk)\, {\cal Y}(\phi_\bp)$ and
  $\lambda>0$ is the dimensionless coupling constant, $\nu_F$ is the density of states at the Fermi level per spin projection, ${\cal Y}(\phi_\bk)=\sqrt{2}\cos2\phi_\bk$ is the normalized $d$-wave form factor, and $\phi_\bk$ is the polar angle of $\bk$ measured from the antinodal $x$ axis as shown in Fig. \ref{Fig1}. Lastly, ${\mathbf A}$ appearing in \eqref{Eq1} is the time dependent vector potential 
\beg\label{VectorPotential}
{\mathbf A}(\br,t)=\left(\frac{c{\mathbf E}}{i\omega}\right)e^{-i\omega t}+\left(-\frac{c{\mathbf E}^*}{i\omega}\right)e^{i\omega t}
\en
considered here at zero photon momentum, $\bq=0$; the electric field $-c^{-1}\partial_t{\mathbf A}=\bE e^{-i\omega t}+\bE^*e^{i\omega t}$ has the complex amplitude $\bE$. 

The main tool of our study will be quasiclassical Green's function defined in Keldysh and Nambu spaces \cite{SereneRainer1983,Kopnin2001,Kamenev2009}:
\beg\label{KeldyshProps}
\check{g}=\left(\begin{matrix} \hat{g}^R & \hat{g}^K \\ 0 & \hat{g}^A\end{matrix}\right).
\en
This function is found by solving the quasiclassical Eilenberger equation \cite{Eilenberger1968,Larkin1965,LarkinVertex,BELZIG1999,LiDzero2024}:
\beg\label{EilenMain}
\begin{aligned}
&\big[\eps\check\tau_3+\check\Delta_\bn(t)-\check\sigma_{\rm imp}(t)\stackrel{\circ}{,}\gc\big]_{-}
+\frac{i}{2}\big[\check\tau_3,\partial_t\gc\big]_{+}\\
&=-\frac{ev_F}{c}\big[(\bn\!\cdot\!{\mathbf A})\check\tau_3\stackrel{\circ}{,}\gc\big]_{-},
\end{aligned}
\en
where $\bn=(\cos\phi_\bn,\sin\phi_\bn)$ is the direction of the Fermi velocity $v_F\bn$, $[\hat{A},\hat{B}]_{\pm}=\hat{A}\hat{B}\pm\hat{B}\hat{A}$, $\check{\tau}_3=\hat{\tau}_3\otimes\hat{1}_{2\times 2}$, $\hat{\tau}_3$ is the third Pauli matrix acting in Nambu space, $\check\sigma_{\rm imp}$ is the self-energy part due to scattering on potential impurities which we will discuss in details below and ${a}_\eps(t)\circ b_\eps(t)$ denotes the Groenewold--Moyal product \cite{Kamenev2011} which in Eq. \eqref{EilenMain} involves time-dependent terms only. The quasiclassical function \eqref{KeldyshProps} satisfies the normalization condition $\gc\circ\gc=\check{\mathbbm 1}$. In Eq. \eqref{EilenMain} $\check{\Delta}_\bn(t)$ is the pairing field matrix and generally its matrix structure is such that it is diagonal in Keldysh space while in Nambu space it is given by the linear superposition of the first and second Pauli matrices $\hat{\tau}_{1,2}$. In this paper we will take the order parameter as  
\beg\label{DefineOP}
\check{\Delta}_\bn(t)=(i\hat{\tau}_2\otimes\hat{1}_{2\times 2}){\cal Y}(\phi_\bn)\Delta(t),
\en
which is exact for a uniform in-plane electromagnetic field ($\bq=0$, normal incidence) and for Born or unitary disorder scattering: under these conditions the matrices $\eps\check\tau_3$, $\check\Delta_\bn$, $\check\sigma_{\rm imp}$ and
$(\bn\cdot{\mathbf A})\check\tau_3$ in Eq.~\eqref{EilenMain} change sign under conjugation by $\check\tau_1$, so the phase
component $i\hat\tau_1\,{\rm Im}\,\Delta$ of the pairing field is never generated. Its amplitude, $\Delta(t)$, is determined self-consistently via
\beg\label{Self4Eli}
\Delta(t)=\frac{\lam}{8}\int\limits_{-\omega_D}^{\omega_D}\!d\eps\,\avg{\Ycal_\bn\Tr\big[(-\itt)\,\hat{g}_{\bn\eps}^K(t)\big]},
\en
where $\avg{\cdots}=\int_0^{2\pi}(d\phi_\bn/2\pi)(\cdots)$ denotes the averaging over the direction of momentum on the Fermi surface, $\Ycal_\bn\equiv{\cal Y}(\phi_\bn)$ with $\avg{\Ycal_\bn^2}=1$, and $\omega_D$ is the energy cutoff of the pairing interaction. In passing we note that \eqref{DefineOP} is adequate here because at finite in-plane momentum the gradient term $v_F(\bn\cdot\nabla)\gc$ drives the phase channel. In two dimensions the corresponding plasma mode, $\omega_{\rm pl}^2=2\pi n_se^2q/m$, is gapless \cite{Kazi2026}, but for a uniform film it lies
outside the light cone and is not excited resonantly by free-space light.

Elastic scattering off point-like impurities with potential $u$ and concentration $n_{\rm imp}$ will be treated in the
self-consistent $T$-matrix approximation \cite{AG1961,AGD,Fominov2011}. Summing the multiple scattering off one impurity and averaging over impurity positions gives:
\beg\label{Tmatrix}
\begin{aligned}
&\check\sigma_{\rm imp}=\Gamma\big(\cot\delta_0+i\avg{\gc}\big)^{-1},
\end{aligned}
\en
where $\Gamma={n_{\rm imp}}/{\pi\nu_F}$, $\cot\delta_0={1}/{\pi\nu_Fu}$, $\delta_0$ is the $s$-wave scattering phase shift and the inverse is taken in Keldysh and Nambu spaces. In the normal state $\avg{\gh^R}=\ttau3$, and \eqref{Tmatrix} gives the single-particle rate $\Gel/2=\Gamma/(1+\cot^2\delta_0)=\Gamma\sin^2\delta_0$, which is used below to compare different scatterers. The two limits are
\beg\label{BornUnitary}
\begin{aligned}
&\check\sigma_{\rm imp}=-\frac{i\Gel}{2}\left\{
\begin{matrix} 
\avg{\gc},\quad |\cot\delta_0|\to\infty,\\ 
\avg{\gc}^{-1},\quad \cot\delta_0=0.
\end{matrix}
\right.
\end{aligned}
\en
In the Born limit $(|\cot\delta_0|\to\infty)$ the constant term $\Gamma\tan\delta_0=n_{\rm imp}u$ has been dropped because it commutes with $\gc$. In what follows we will focus on the case of Born scattering. 

\subsection{Ground state}
In equilibrium we look for the solution of \eqref{EilenMain} in the form
\beg\label{gRAGS}
\gh^{R(A)}_{\bn\eps}=\ttau3\,g^{R(A)}_{\bn\eps}+\itt\,f^{R(A)}_{\bn\eps} 
\en
and $\gh^K_{\bn\eps}=(\gh^R_{\bn\eps}-\gh^A_{\bn\eps})\,t_\eps$ with $t_\eps=\tanh(\eps/2T)$. For brevity let us denote this solution by $\gc_0$. For $d$-wave pairing $f_{\bn\eps}$ is odd in $\Ycal_\bn$, so $\avg{f_{\bn\eps}}=0$ which implies that $\avg{\gc_0}$ must be proportional to $\check{\tau}_3$. It then obtains
\beg\label{Avbn}
\avg{\hat{g}_{\bn\eps}^{R(A)}}=\ttau3\Gcal^{R(A)}_\eps.
\en

The inverse in \eqref{Tmatrix} is then trivial since in the ground state its $\ttau0$
part commutes with $\gc_0$, and its $\ttau3$ part renormalizes the energy:
\beg\label{xitilde}
\tveps^{R}_\eps=\eps+\Phi[\Gcal^R_\eps],\quad \tveps^{A}_\eps=(\tveps^R_\eps)^*,
\en
where 
\beg\label{PhiDef}
\begin{aligned}
&\Phi[\Gcal]=\frac{i\Gamma\Gcal}{\cot^2\delta_0+\Gcal^2}.
\end{aligned}
\en
It then follows that
\beg\label{gnRA}
g^{R(A)}_{\bn\eps}=\frac{\tveps^{R(A)}_\eps}{\eta^{R(A)}_{\bn\eps}}, \quad f^{R(A)}_{\bn\eps}=\frac{\Delta_\bn}{\eta^{R(A)}_{\bn\eps}},
\en
where $\eta^R_{\bn\eps}=\tveps^R_\eps\sqrt{1-(\Delta_\bn/\tveps^R_\eps)^2}$, where $\Delta_\bn={\cal Y}(\phi_\bn)\Delta$, on the branch $\Im\left(\eta^R_{\bn\eps}\right)>0$, $\eta^A_{\bn\eps}=-[\eta^R_{\bn\eps}]^*$. We can now insert \eqref{gnRA} into \eqref{Avbn} to derive the self-consistency equation for the function
\beg\label{ComputeGcalR}
\Gcal^R_\eps=\frac{2}{\pi}K\left(\frac{\sqrt2\Delta}{\tveps^R_\eps}\right),
\en
where $K(k)=\int_0^{\pi/2}d\vartheta/\sqrt{1-k^2\sin^2\vartheta}$ is the complete elliptic integral
of the first kind, continued to complex $k$ with the branch fixed by $\Gcal^R_\eps\to1$ for $|\eps|\to\infty$. Thus, equations (\ref{Tmatrix},\ref{Avbn},\ref{ComputeGcalR}) fully determine the retarded and advanced parts of $\check\sigma_{\rm imp}$ in equilibrium. In turn its Keldysh part is given by the usual parametrization $\sigma^K=(\hat\sigma^R-\hat\sigma^A)t_\eps$ for any $\delta_0$. 

\begin{figure}
\includegraphics[width=\linewidth]{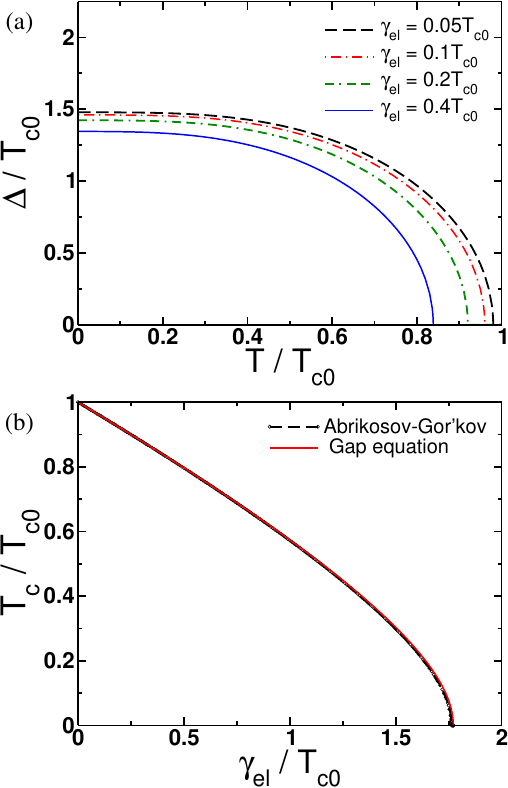}
\caption{Equilibrium state with Born disorder. (a) Amplitude $\Delta(T)$ of the order parameter
$\Delta_\bn=\Ycal(\phi_\bn)\Delta$ for $\Gel/T_{c0}=0.05$, $0.1$, $0.2$ and $0.4$, the values used in
Figs.~\ref{Fig3} and~\ref{Fig4}. It is obtained from the gap equation~\eqref{Self4Eli} with the cutoff
$\omega_D=9.02\,T_{c0}$ and the self-consistent energy $\tveps^R_\eps$ of Eqs.~\eqref{xitilde} and~\eqref{ComputeGcalR}
in the Born limit. (b) Critical temperature as a function of $\Gel$: the Abrikosov--Gor'kov form
$\ln(T_{c0}/T_c)=\psi(1/2+\Gel/4\pi T_c)-\psi(1/2)$, with $\psi$ the digamma function (dashed), and the linearized
gap equation~\eqref{Self4Eli} (solid). The two coincide for $\omega_D\gg T_{c0}$ and here differ visibly only near the
critical value, which is $\Gel\approx1.764\,T_{c0}$ for the Abrikosov--Gor'kov form and $1.772\,T_{c0}$ with the
finite cutoff.}
\label{Fig2}
\end{figure}
In Fig. \ref{Fig2}(a) we show the temperature dependence of the order parameter for different values of disorder scattering rate. As one may have expected, the critical temperature shown in Fig. \ref{Fig2}(b) decreases with the increase in $\gamma_{\rm el}$. This dependence computed using the self-consistency equation \eqref{Self4Eli} is very well approximated by the celebrated Abrikosov-Gor'kov curve for the critical temperature of $s$-wave superconductors with paramagnetic impurities \cite{AG1961}. We will use these results in our analysis of the third harmonic generation below.

\subsection{Outline of the perturbative calculation}\label{sec:scheme}
In order to compute the nonlinear corrections to the current density, we will develop the perturbative expansion
\beg\label{gexpand}
\gc=\gc_0+\gc_1+\gc_2+\gc_3+\dots
\en
with $\gc_m=O({\mathbf A}^m)$ and solve \eqref{EilenMain} at each order. As we will show below the components needed
  for the third harmonic at order ${\mathbf A}^3$ are $\gc_1\propto e^{-i\omega t}$, $\gc_2\propto e^{-2i\omega t}$ and
  $\gc_3\propto e^{-3i\omega t}$. The static component of $\gc_2$ will not be needed because when combined with $\gc_1$ it produces the harmonic $\omega$, not $3\omega$. It is also useful to keep in mind that by parity $\bn\to-\bn$ it follows that  $\avg{\gc_1}=\avg{\gc_3}=0$. Therefore, the only disorder vertex correction entering the third harmonic
is $\langle{\gc_2^{(2\omega)}}\rangle_\bn$. 

Similarly, the order parameter \eqref{DefineOP} will also be modified, $\Delta(t)=\Delta+\delta\Delta_2(t)$. While its first order correction is zero, its second order correction $\delta\Delta_2(t)$ will be proportional to the longitudinal pair susceptibility \cite{Kazi2026} and will ultimately contribute to the third harmonic response of the current density.

\section{Second order corrections}\label{sec:2w}
Calculation of the first order corrections to $\check{g}$ is described in Appendix \ref{AppendixB:first_order} so here we start with the calculation of the second order corrections which determine the variation of the order parameter. 
The second order correction to the function $\check{g}$ is formally given by 
\beg\label{g2RA}
\begin{aligned}
\check{g}_{2}(\bn\eps;t)&=\check{g}_{2}(\bn\eps;\omega)+\check{g}_{2,+}(\bn\eps;\omega)e^{-2i\omega t}\\&+\check{g}_{2,-}(\bn\eps;\omega)e^{2i\omega t}.
\end{aligned}
\en
As we have mentioned above for the third harmonic calculation we will be needing the second term only, so we discard the first and the third terms and re-define $\check{g}_{2,+}\to \check{g}_2$. Likewise, for the longitudinal component of the pairing field  we write
\beg\label{deltaDelta}
\begin{aligned}
\delta\hat{\Delta}_\bn(\br,t)=i\hat{\tau}_2\delta\Delta_\bn(\omega)e^{-2i\omega t}.
\end{aligned}
\en
For $d$-wave pairing $\delta\Delta_\bn(\omega)=\Ycal_\bn\,\delta\Delta_{2\omega}$, and $\delta\check\Delta_\bn$ denotes the
corresponding matrix $(\itt\otimes\hat1_{2\times2})\Ycal_\bn\delta\Delta_{2\omega}$, diagonal in Keldysh space.

\subsection{Second order corrections to a quasiclassical propagator}
For components oscillating as $e^{-ia\omega t}$ the Groenewold--Moyal product reduces to the shift rule
\beg\label{GM_Shift}
\hat X_{\eps}e^{-ia\omega t}\circ\hat Y_{\eps}e^{-ib\omega t}=\hat X_{\eps+b\frac{\omega}{2}}\,\hat Y_{\eps-a\frac{\omega}{2}}
e^{-i(a+b)\omega t}. 
\en
Collecting the terms of second order in
${\mathbf A}$ that oscillate as $e^{-2i\omega t}$ in Eq.~\eqref{EilenMain} we find
\beg\label{Eq2w}
\begin{aligned}
&\check H_{\eps+\omega}\,\gc_2-\gc_2\,\check H_{\eps-\omega}+\big(\delta\check\Delta_\bn\,\gc_{\bn\eps-\omega}-\gc_{\bn\eps+\omega}\,\delta\check\Delta_\bn\big)\\
&={\cal A}_+\Big[\gc_1\big(\bn,\eps+\tfrac\omega2\big)\check\tau_3-\check\tau_3\,\gc_1\big(\bn,\eps-\tfrac\omega2\big)\Big]\\
&+\big(\delta\check\sigma\,\gc_{\bn\eps-\omega}-\gc_{\bn\eps+\omega}\,\delta\check\sigma\big).
\end{aligned}
\en
Here $\gc_2\equiv\gc_2(\bn\eps;\omega)$, $\gc_{\bn\eps}\equiv\gc_0(\bn\eps)$ is the equilibrium function, Eq.~\eqref{gRAGS},
$\check H_\eps=\eps\check\tau_3+\check\Delta_\bn-\check\sigma_{\rm imp}(\eps)$ contains the equilibrium self-energy,
$\gc_1(\bn,\eps)\equiv\gc_{1,+}(\bn\eps;\omega)$ is the first-order function of Appendix~\ref{AppendixB:first_order}, and
\beg\label{Aplus}
{\cal A}_+=\left(\frac{ev_F}{i\omega}\right)(\bn\cdot\bE).
\en
The last term of Eq. \eqref{Eq2w} contains the second-order impurity self-energy, which in the Born
limit of Eq.~\eqref{BornUnitary} is $\delta\check\sigma=-(i\Gel/2)\avg{\gc_2}$. There are no first-order corrections to the
disorder self-energy and to the pairing field. The normalization condition at this order reads
\beg\label{Norm2w}
\begin{aligned}
&\gc_{\bn\eps+\omega}\,\gc_2+\gc_2\,\gc_{\bn\eps-\omega}
=-\gc_1\big(\bn,\eps+\tfrac\omega2\big)\,\gc_1\big(\bn,\eps-\tfrac\omega2\big).
\end{aligned}
\en

Similar to how it was done for $\hat{g}_1^K$,  we write the Keldysh component in the form
\beg\label{g2Kansatz2w}
\gh_2^K=\gh_2^R\,t_{\eps-\omega}-t_{\eps+\omega}\,\gh_2^A+\delta\gh_2^K .
\en
The resulting expressions for the functions $\gh_2^{R(A)}$ and $\delta\gh_2^K$ are given in Appendix \ref{AppendixC}. 

\begin{figure}
\includegraphics[width=0.850\linewidth]{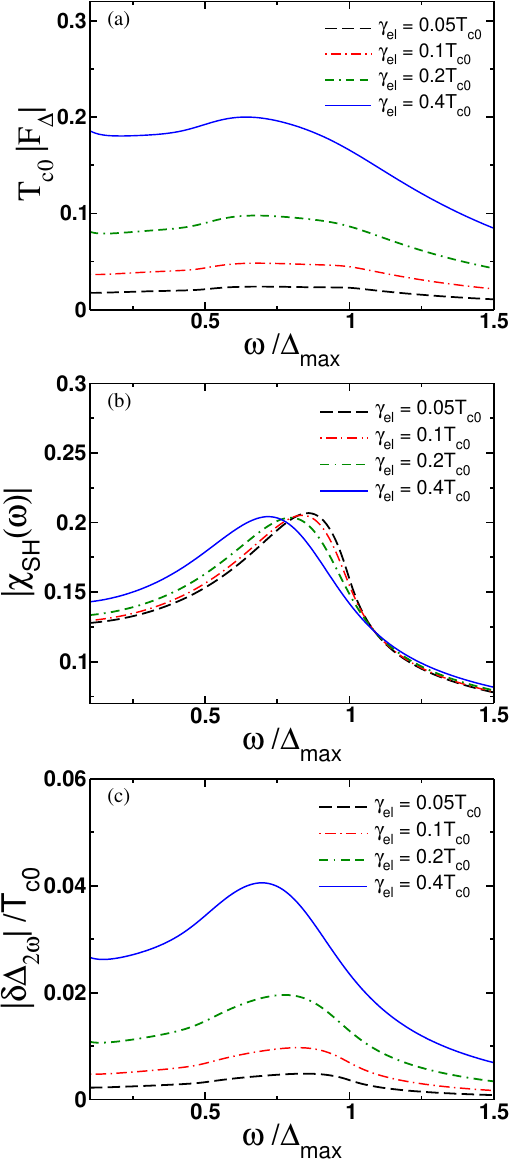}
\caption{Second-order response at $2\omega$ for Born disorder as a function of the drive frequency, $\omega/\Delta_{\max}$ with
$\Delta_{\max}=\sqrt2\,\Delta(T,\Gel)$, at $T=0.2\,T_{c0}$ and $\Gel/T_{c0}=0.05$, $0.1$, $0.2$, $0.4$.
Panel (a): Source of the amplitude oscillation, $|F_\Delta|$ of Eq.~\eqref{FDelta2w}, in units of $1/T_{c0}$. Panel (b): Longitudinal pair susceptibility $|\chi_{\rm SH}|$, Eq.~\eqref{chiSH2w}.
Panel (c): Amplitude of the gap oscillation, $|\delta\Delta_{2\omega}|$ of Eq.~\eqref{Eq4dDelta2w}, for linear polarization, in units of
$T_{c0}(ev_FE/\omega T_{c0})^2$.}
\label{Fig3}
\end{figure}

\subsection{Second order correction to the pairing field}
The Fourier component of the order parameter amplitude at the frequency $2\omega$, Eq.~\eqref{Self4Eli}, gives
\beg\label{deltaDelta2w}
\delta\Delta_{2\omega}=\frac{\lam}{8}\int d\eps\,\avg{\Ycal_\bn\Tr[(-\itt)\gh_2^K]}
\en
with $\gh_2^K$ of Eq.~\eqref{g2Kansatz2w}.
Inserting Eqs.~(\ref{Solution2w})--(\ref{ClosureX2}) and eliminating $8/\lam$ with the equilibrium gap equation,
$8/\lam=(2/\Delta)\int d\eps\,(\Fcal^R_\eps-\Fcal^A_\eps)\,t_\eps$, we obtain
\beg\label{Eq4dDelta2w}
\chi_{\rm SH}^{-1}(\omega)\,\delta\Delta_{2\omega}=\Big(\frac{ev_F}{\omega}\Big)^{\!2}(\bE\cdot\bE)\,F_\Delta(\omega),
\en
where $\chi_{\rm SH}(\omega)$ is the longitudinal pair susceptibility of the $d$-wave superconductor \cite{Awelewa2026,Kazi2026},
which is dressed by disorder including the vertex term $\propto\Fcal[a_+,b_-]^2$:
\beg\label{chiSH2w}
\begin{aligned}
\chi_{\rm SH}^{-1}(\omega)&=2\int d\eps\,\bigg\{\frac{\Fcal^R_\eps-\Fcal^A_\eps}{\Delta}\,t_\eps\\
&-\sum_{ab}c_{ab}w_{ab}\bigg[\avg{\frac{\Ycal_\bn^2\bar B_{ab}}{\eta^a_++\eta^b_-}}\\
&-\frac{i\Gel}{2}\,\frac{\tveps^a_+-\tveps^b_-}{2\omega}\,\Fcal[a_+,b_-]^2\bigg]\bigg\}.
\end{aligned}
\en
In Fig. \ref{Fig3}b,c we plot the absolute value of these functions for different values of disorder scattering rate. It important to emphasize that these quantities scale differently with $\gamma_{\rm el}$:  while longitudinal pair susceptibility remains varies within the same range for different $\gamma_{\rm el}$, the variation of the pairing amplitude $\delta\Delta_{2\omega}\sim\gamma_{\rm el}$. This indeed confirms that the excitation of Higgs mode in $d$-wave superconductors is promoted by elastic scattering just like it happens for the $s$-wave pairing \cite{Silaev2019}.

Function $F_\Delta(\omega)$ in \eqref{Eq4dDelta2w} accounts for an external drive:
\beg\label{FDelta2w}
\begin{aligned}
F_\Delta(\omega)&=-\int d\eps\sum_{ab}c_{ab}\bigg[\Fcal^{(3)}_{ab}\\
&\quad+\frac{i\Gel}{2}\,\left(\frac{\tveps^a_+-\tveps^b_-}{2\omega}\right)\,\Fcal[a_+,b_-]\,\Gcal^{(3)}_{ab}\bigg],
\end{aligned}
\en
where $c_{RR}=t_{\eps-\omega}$, $c_{AA}=-t_{\eps+\omega}$ and $c_{RA}=1$. Both energy integrals converge, so that the cutoff
$\omega_D$ has been sent to infinity (Appendix~\ref{AppendixC}).  The source requires only the two Fermi-surface functions $\Gcal^a_\eps$ and $\Fcal^a_\eps$ at
the six points $\tveps^a_s$. In the clean limit the divided differences become finite differences
\beg\label{CleanLimitSum}
\sum_{ab}c_{ab}\Fcal^{(3)}_{ab}\to\frac{K_{\eps+\omega}-2K_\eps+K_{\eps-\omega}}{2\omega^2}
\en
with $K_\eps=(\Fcal^R_\eps-\Fcal^A_\eps)t_\eps$. The energy integral of this second difference vanishes, and the vertex term of Eq.~\eqref{FDelta2w} is proportional to
$\Gel$. Hence $F_\Delta\to0$ for $\Gel\to0$: in the clean limit the field does not drive the amplitude mode at $2\omega$,
which is in full agreement with Eq.~\eqref{FiniteDifferences} in Appendix \ref{AppendixA:gauge}. At finite $\Gel$ two contributions appear, one from the displacement of the
points $\tveps^a_s$ from $\eps+s\omega$ and one from the disorder vertex.  In Fig. \ref{Fig3}a we plot $|F_\Delta(\omega)|$ for different values of $\gamma_{\rm el}$ which indeed shows that at a given frequency this function increases with increasing scattering rate. 

\section{Third-harmonic generation}\label{sec:thg}
In this Section we derive an expression for the third harmonic component of the current density. By construction, the resulting expressions appear to be complicated. For this reason here we give only resulting expressions in terms of the some generic functions of our theory. The explicit expressions for these functions along with the derivation details can be found in Appendix \ref{app:third}.
\subsection{Third-order equations}
Collecting the terms of third order in ${\mathbf A}$ that oscillate as $e^{-3i\omega t}$ in Eq.~\eqref{EilenMain} and
using the shift rule \eqref{GM_Shift}, we find
\beg\label{Eq3w}
\begin{aligned}
&\check H_{\eps+\frac{3\omega}2}\,\gc_3-\gc_3\,\check H_{\eps-\frac{3\omega}2}+\big[\delta\check\Delta_\bn\stackrel{\circ},\gc_1\big](\eps)\\
&={\cal A}_+\Big[\gc_2\big(\bn,\eps+\tfrac\omega2\big)\check\tau_3-\check\tau_3\,\gc_2\big(\bn,\eps-\tfrac\omega2\big)\Big]\\
&+\big[\delta\check\sigma_2\stackrel{\circ},\gc_1\big](\eps).
\end{aligned}
\en
Here $\gc_3\equiv\gc_{3,+}(\bn\eps;\omega)$, $\gc_2(\bn,\eps)\equiv\gc_2(\bn\eps;\omega)$ is the solution of
Sec.~\ref{sec:2w}, $\delta\check\sigma_2(\eps)=-(i\Gel/2)\avg{\gc_2(\bn,\eps)}$ is the self-energy $\delta\check\sigma$ of
Eq.~\eqref{Eq2w} at the centre energy $\eps$, and
\beg\label{VertexSource3}
\begin{aligned}
\big[\delta\check\sigma_2\stackrel{\circ},\gc_{1}\big](\eps)&=\delta\check\sigma_2\big(\eps+\tfrac\omega2\big)\,\gc_{1}(\bn,\eps-\omega)\\
&-\gc_{1}(\bn,\eps+\omega)\,\delta\check\sigma_2\big(\eps-\tfrac\omega2\big).
\end{aligned}
\en
The commutator with $\delta\check\Delta_\bn$ is defined in the same way, with the energy-independent matrix
$\delta\check\Delta_\bn$ in place of $\delta\check\sigma_2$. The normalization condition at this order reads
\beg\label{Norm3w}
\begin{aligned}
&\gc_{\bn\eps+\frac{3\omega}2}\,\gc_3+\gc_3\,\gc_{\bn\eps-\frac{3\omega}2}=\\&-\gc_1(\bn,\eps+\omega)\,\gc_2\big(\bn,\eps-\tfrac\omega2\big)\\&-\gc_2\big(\bn,\eps+\tfrac\omega2\big)\,\gc_1(\bn,\eps-\omega).
\end{aligned}
\en
Since $\gc_3$ is odd under $\bn\to-\bn$, both $\avg{\gc_3}$ and $\avg{\Ycal_\bn\gc_3}$ vanish.
There is therefore no third-order self-energy and no correction to the pairing field at $3\omega$. The collective
quantities of Sec.~\ref{sec:2w}, $\delta\Delta_{2\omega}$ and $\delta\check\sigma_2$, enter Eq.~\eqref{Eq3w} only through
their commutators with $\gc_1$.

As in Eq.~\eqref{g2Kansatz2w} we write the Keldysh component in the form
\beg\label{g3Kansatz}
\gh_3^K=\gh_3^R\,t_{\eps-\frac{3\omega}2}-t_{\eps+\frac{3\omega}2}\,\gh_3^A+\delta\gh_3^K .
\en
Equations (\ref{Eq3w},\ref{Norm3w}) then split into three problems for the matrices $\Xi_{RR}=\gh_3^R$,
$\Xi_{AA}=\gh_3^A$ and $\Xi_{RA}=\delta\gh_3^K$. They have the form solved in Appendix~\ref{AppendixC},
Eq.~\eqref{ClosedForm2w}, with the points $a_{3/2}$ and $b_{-3/2}$ in place of $a_+$ and $b_-$, and their sources are built
from the first- and second-order functions (Appendix~\ref{app:third}). 

Equation~\eqref{Eq3w} shows how the third harmonic is generated. A component oscillating as $e^{-3i\omega t}$ requires the
$2\omega$ state and one more photon, because the static part of $\gc_2$ combined with $\gc_1$ gives the first harmonic only.
The $2\omega$ state consists of the quasiparticle response $\gc_2$ and of the two collective fields it generates, the
oscillation of the gap amplitude $\delta\Delta_{2\omega}$ and the impurity self-energy $\delta\check\sigma_2$. The third
photon either acts on $\gc_2$, which is the term proportional to ${\cal A}_+$ in Eq.~\eqref{Eq3w}, or produces the
first-order response $\gc_1$ on which the collective fields then act, which gives the two commutators. Since $\gc_3$ is
odd in $\bn$, no collective field is generated at $3\omega$, and this is why the third harmonic is the single-particle read-out of the
$2\omega$ state.

The solution of these equations, Eqs.~(\ref{DriveDD3}) and (\ref{Solution3w}) of Appendix~\ref{app:third}, is built from
divided differences over the complex energies $\tveps^a_s$ with $s=\pm\frac12,\pm\frac32$: the third divided difference
of $\gh$ for the direct three-photon process, and divided differences of $V$ for the processes that involve the $2\omega$
fields. The denominator of a divided difference,
\beg
\tveps^a_s-\tveps^b_{s'}=(s-s')\,\omega+\frac{i\Gel}{2}\big(\Gcal^a_{\eps+s\omega}-\Gcal^b_{\eps+s'\omega}\big),
\en
is the photon energy absorbed between two quasiparticle states, corrected by the difference of their impurity
self-energies. In the Keldysh sector (i.e. $ab=RA$), which describes the change of the quasiparticle distribution, each step from the
retarded to the advanced branch carries the difference of the occupation factors $t_s$ of the two states it connects, and
the denominator of this step has the imaginary part $\Im\tveps^R_s+\Im\tveps^R_{s'}$, the sum of the scattering rates of
the two states, as for a real transition between them. In the clean limit all points move to $\eps+s\omega$, the divided
differences become finite differences with step $\omega$, and the energy integral of a finite difference is fixed by the
values of the function at $\eps\to\pm\infty$. As we will show below for the combinations that enter the current this integral vanishes. 
This is the gauge invariance argument that we developed in Appendix~\ref{AppendixA:gauge}: on each trajectory a uniform field only
shifts the quasiparticle energies, and the shift has no observable consequence. Disorder lifts the cancellation in two
ways. It displaces the points $\tveps^a_s$ from $\eps+s\omega$ by amounts that depend on the energy and on the branch, and
it produces the $2\omega$ self-energy $\delta\check\sigma_2$. Both effects originate in the angular average of the impurity
self-energy, that is, in momentum relaxation. The linear response behaves in the same way: without momentum relaxation it
reduces to the London term (Appendix~\ref{AppendixA:gauge}), and a uniform field is not absorbed.

\subsection{Higgs and quasiparticle contributions}
Within the quasiclassical formalism the current density is
\beg\label{MainCurrent}
\mathbf j(t)=-\frac{e\nu_Fv_F}{4}\int_{-\infty}^{\infty}\!d\eps\,\avg{\bn\,\Tr\big[\ttau3\hat g^K(\bn\eps;t)\big]}.
\en
We can now use this formula along with the results for the third order correction to the $\hat{g}^K$ to evaluate the
third harmonic contribution to the current density. With Eq.~\eqref{g3Kansatz} the amplitude of the component
$e^{-3i\omega t}$ is
\beg\label{Current3w}
\mathbf j^{(3\omega)}=-\frac{e\nu_Fv_F}{2}\int d\eps\sum_{ab}c_{ab}\avg{\bn\,[\Xi_{ab}]_{\ttau3}},
\en
where $[\hat X]_{\ttau3}=\frac12\Tr[\ttau3\hat X]$ and, at this order, $c_{RR}=t_{\eps-3\omega/2}$, $c_{AA}=-t_{\eps+3\omega/2}$ and
$c_{RA}=1$ (see Appendix \ref{app:third}). Then
\beg\label{Trace3w}
[\Xi_{ab}]_{\ttau3}={\cal A}_+^3\,g^{(4)}_{ab}+{\cal A}_+\Big({\cal S}_{ab}\big[u^{(3)},g\big]-{\cal S}_{ab}\big[u^{(2)},f\big]\Big),
\en
where the functions appearing in the right hand side have been defined in Appendix \ref{app:third} and $g^{(4)}_{ab}$ are the combinations of Eq.~\eqref{DriveDD3} for $g$.

The Fermi-surface average in Eq.~\eqref{Current3w} involves one or three powers of $\bn\cdot\bE$. For any function $h$ of
$\Ycal_\bn^2$
\beg\label{PolAverages3}
\begin{aligned}
&\avg{\bn(\bn\cdot\bE)h}=\frac{\bE}{2}\avg{h},\\
&\avg{\bn(\bn\cdot\bE)\Ycal_\bn h}=\frac{\bE'}{2\sqrt2}\avg{\Ycal_\bn^2h},\\
&\avg{\bn(\bn\cdot\bE)^3h}=\frac{3\avg{h}-\avg{h\cos4\phi_\bn}}{8}\,(\bE\cdot\bE)\bE\\
&\qquad+\frac{\avg{h\cos4\phi_\bn}}{4}\,(E_x^2-E_y^2)\bE',
\end{aligned}
\en
where $\bE'\equiv(E_x,-E_y)$. Since $g$ and $\Ycal_\bn f$ are functions of $\Ycal_\bn^2$, while $f$ is odd in $\Ycal_\bn$,
the averages involve $\Gcal^a_\eps$, $\Fcal^a_\eps$ and, in the drive term only, the fourth harmonic
\beg\label{KcalDef}
{\cal K}^a_\eps=\avg{\cos4\phi_\bn\,g^a_{\bn\eps}}.
\en
The current is therefore determined by the three Fermi-surface functions $\Gcal$, $\Fcal$ and ${\cal K}$ at the eight
points $\tveps^a_s$, together with the $2\omega$ amplitudes of Sec.~\ref{sec:2w}. We write the latter as
\beg\label{VertexAmps}
\begin{aligned}
\frac{i\Gel}{2}X^{(3)}_{ab}&=-\Big(\frac{ev_F}{\omega}\Big)^{\!2}\frac{\bE\cdot\bE}{2}\,\varrho_{ab}-r_{ab}\,\delta\Delta_{2\omega},\\
\frac{i\Gel}{2}X^{(2)}_{ab}&=-\Big(\frac{ev_F}{\omega}\Big)^{\!2}\left(\frac{E_x^2-E_y^2}{2\sqrt2}\right)\,\psi_{ab},
\end{aligned}
\en
where, by Eqs.~(\ref{WardClosure},\ref{ClosureX2}), we defined
\beg\label{VertexAmps2}
\begin{aligned}
&\varrho_{ab}=\frac{i\Gel}{2}\,\left(\frac{\tveps^a_+-\tveps^b_-}{2\omega}\right)\,\Gcal^{(3)}_{ab},\qquad
\psi_{ab}=\frac{i\Gel}{2}\,\frac{\Fcal^{(3)}_{ab}}{\kappa'_{ab}},\\
&r_{ab}=\frac{i\Gel}{2}\,\left(\frac{\tveps^a_+-\tveps^b_-}{2\omega}\right)\,w_{ab}\Fcal[a_+,b_-].
\end{aligned}
\en
The two vertex amplitudes \eqref{VertexAmps} have a simple symmetry content. Under the point group of the square lattice the bilinear
$E_iE_j$ splits into the $A_{1g}$ combination $E_x^2+E_y^2$, the $B_{1g}$ combination $E_x^2-E_y^2$ and the $B_{2g}$
combination $2E_xE_y$. The amplitude mode changes the magnitude of the order parameter but not its form factor, and it is
therefore driven by the $A_{1g}$ combination only, $\delta\Delta_{2\omega}\propto\bE\cdot\bE$ in Eq.~\eqref{Eq4dDelta2w}.
The $\ttau3$ component $X^{(3)}$ modulates the energy renormalization $\eps\to\tveps$ at $2\omega$. It is isotropic as well
and is driven by $\bE\cdot\bE$, directly and through $\delta\Delta_{2\omega}$. The $\itt$ component $X^{(2)}$ is an isotropic
anomalous self-energy, that is, an $s$-wave anomalous component produced by impurity scattering and oscillating at
$2\omega$. It is absent in equilibrium, where
$\avg{f_{\bn\eps}}=0$, and is produced by the $B_{1g}$ combination of the field, because the $d_{x^2-y^2}$ form factor
$\Ycal_\bn$ itself belongs to $B_{1g}$ and $B_{1g}\otimes B_{1g}=A_{1g}$. The $B_{2g}$ combination does not couple, since
all equilibrium functions are even under $\phi_\bn\to-\phi_\bn$. For circular polarization $\bE\cdot\bE=0$: the amplitude
mode and the $\ttau3$ vertex are not excited, and the $B_{1g}$ channel is the only collective response at $2\omega$.

We define the Higgs contribution $\mathbf j_{\rm H}$ as the part of $\mathbf j^{(3\omega)}$ that is proportional to
$\delta\Delta_{2\omega}$, including the part of $\delta\hat\sigma_{ab}$ driven by $\delta\Delta_{2\omega}$ (the term
$r_{ab}$), and the quasiparticle contribution $\mathbf j_{\rm qp}$ as the rest,
$\mathbf j^{(3\omega)}=\mathbf j_{\rm H}+\mathbf j_{\rm qp}$. Inserting
Eqs.~(\ref{Trace3w})--(\ref{VertexAmps2}) into Eq.~\eqref{Current3w} we find
\beg\label{jHiggs}
\mathbf j_{\rm H}=\frac{e\nu_Fv_F}{4}\,\frac{ev_F}{i\omega}\,\Lambda_{\rm H}(\omega)\,\delta\Delta_{2\omega}\,\bE,
\en
where function $\Lambda_H$ is given by 
\beg\label{LambdaH}
\Lambda_{\rm H}(\omega)=\int d\eps\sum_{ab}c_{ab}\Big({\cal S}_{ab}[w,\Fcal]+{\cal S}_{ab}[r,\Gcal]\Big),
\en
Taking into account Eq.~\eqref{Eq4dDelta2w}, for the Higgs part of the current we have
$\mathbf j_{\rm H}=j_0\Lambda_{\rm H}(\omega)\chi_{\rm SH}(2\omega)F_\Delta(\omega)(\bE\cdot\bE)\bE$. 

In turn, the quasiparticle part of the current is defined by
\beg\label{jqp3w}
\begin{aligned}
\mathbf j_{\rm qp}&=j_0\int d\eps\sum_{ab}c_{ab}\bigg\{\Big[\frac{3\Gcal^{(4)}_{ab}-{\cal K}^{(4)}_{ab}}{4}\\
&+\frac12{\cal S}_{ab}[\varrho,\Gcal]\Big](\bE\cdot\bE)\bE+\Big[\frac12{\cal K}^{(4)}_{ab}\\
&-\frac14{\cal S}_{ab}[\psi,\Fcal]\Big](E_x^2-E_y^2)\bE'\bigg\},
\end{aligned}
\en
with $j_0=-(e\nu_Fv_F/4)(ev_F/i\omega)^3$. Here $\Gcal^{(4)}_{ab}$ and ${\cal K}^{(4)}_{ab}$ are the combinations of
Eq.~\eqref{DriveDD3} for $\Gcal$ and ${\cal K}$. The coupling $\Lambda_{\rm H}$ of the amplitude oscillation to the
third-harmonic current has the same structure as the source $F_\Delta$ of Eq.~\eqref{FDelta2w}: the direct term contains
$\Fcal$, and the vertex term contains the product of $\Fcal[a_+,b_-]$ with $\Gcal$. 
As we have checked all energy integrals which define the functions in this expression are ultraviolet convergent (Appendix~\ref{app:third}).

We would like to emphasize that both contributions vanish in the clean limit $\gamma_{\rm el}=0$. 
The drive term of $\mathbf j_{\rm qp}$ becomes the energy integral of the
third finite difference of $\gh^K$ (Appendix~\ref{AppendixA:gauge}). This integral vanishes after a shift of the
integration variable, the constant large-$|\eps|$ limit of $\gh^K$ included, because the coefficients $1,-3,3,-1$ of the
third difference multiplied by the shifts $3/2,1/2,-1/2,-3/2$ add up to zero. The vertex terms of $\mathbf j_{\rm qp}$ are
proportional to $\Gel$. In $\mathbf j_{\rm H}$ the coupling $\Lambda_{\rm H}$ vanishes because the $\delta\Delta_{2\omega}$
part of $\gc_3$ becomes the first finite difference of a function that decays at large $|\eps|$, and in addition
$\delta\Delta_{2\omega}\to0$ because $F_\Delta\to0$ (Sec.~\ref{sec:2w}). At finite $\Gel$ the Higgs contribution therefore
contains two factors that vanish in the clean limit, $\Lambda_{\rm H}$ and $F_\Delta$. 

\begin{figure}
\includegraphics[width=0.850\linewidth]{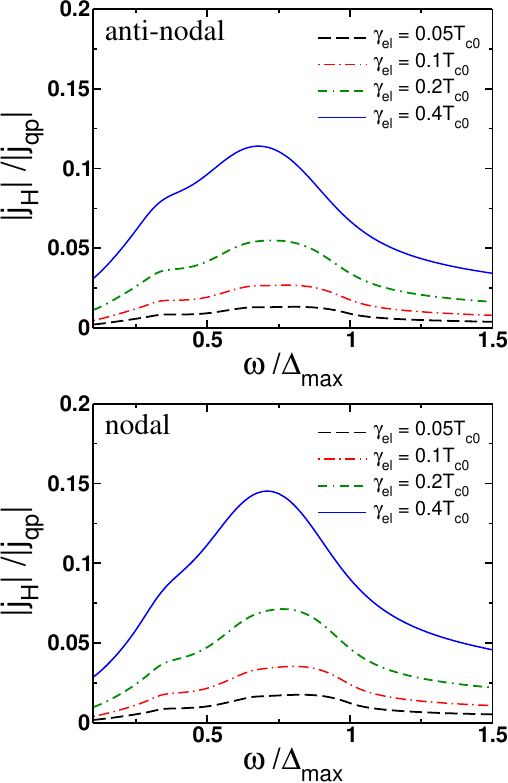}
\caption{Ratio $|\mathbf j_{\rm H}|/|\mathbf j_{\rm qp}|$ of the third-harmonic Higgs and quasiparticle currents for linear polarization of the electric field as a function of $\omega/\Delta_{\max}$, at $T=0.2\,T_{c0}$ and
$\Gel/T_{c0}=0.05$, $0.1$, $0.2$, $0.4$. Top panel shows the result for the field in the anti-nodal direction ($\theta=0$) and the bottom panel shows the results for the field in the nodal direction ($\theta=\pi/4$).}
\label{Fig4}
\end{figure}

This power counting is important as it actually decides which contribution dominates at weak disorder. In Fig. \ref{Fig4} we present the results for the frequency dependence of the ratio $|j_H|/|j_{\rm qp}|$ in the nodal and antinodal direction of the external field. The quasiparticle current requires momentum
relaxation once, whereas the Higgs current requires it twice: once to excite the amplitude oscillation, through $F_\Delta$,
and once to convert it into a current, through $\Lambda_{\rm H}$. As we show in Fig.~\ref{Fig4} our numerical calculation using 
Eqs.~(\ref{chiSH2w}), (\ref{FDelta2w}), (\ref{LambdaH}) and (\ref{AB3w}) confirms this. At fixed $\omega$ and $T$, $F_\Delta$, $\Lambda_{\rm H}$ and the quasiparticle parts of $A$ and $B$ are linear
in $\Gel$ for $\Gel\to0$, while $\chi_{\rm SH}$ has a finite limit, so that $\mathbf j_{\rm H}\propto\Gel^2$ and
$\mathbf j_{\rm qp}\propto\Gel$. The susceptibility does not compensate the additional power of $\Gel$. In the $d$-wave
state the amplitude oscillation decays into pairs of nodal quasiparticles at any frequency, $\Im\chi_{\rm SH}^{-1}\neq0$ for
all $\omega>0$ even for $\Gel,T\to0$, and $|\chi_{\rm SH}|$ has only a broad maximum below $\omega=\Delta_{\max}$ instead of
a resonance at $2\omega=2\Delta_{\max}$. For linear polarization along the nodal or the antinodal direction with $\Gel\le0.4\,T_{c0}$,
$0.1\le\omega/\Delta_{\max}\le1.5$ and $T/T_{c0}=0.2$ and $0.6$, we find for the ratio 
\beg\label{jH2jqp}
\frac{|\mathbf j_{\rm H}|}{|\mathbf j_{\rm qp}|}\lesssim\frac{\Gel}{2\Delta}
\en
with $\Gel/2\Delta$ being a value it reaches near the maximum of $|\chi_{\rm SH}|$. Within the present model the third harmonic in $d$-wave superconductors is therefore dominated by the quasiparticle contribution, and the Higgs contribution is a correction of order $\Gel/\Delta$.

Finally, we note that the disorder vertex is essential for the size of $\mathbf j_{\rm qp}$. For $\Gel\to0$ we find numerically that the energy
integral of the $\ttau3$ vertex term of $A$, $\frac12{\cal S}_{ab}[\varrho,\Gcal]$, is twice that of the drive term
$\propto\Gcal^{(4)}_{ab}$, although the vertex integrand is small at every energy and the drive integrand is not. A
calculation in which disorder enters only through $\tveps$ keeps the drive term alone and, in the range given above,
underestimates the quasiparticle part of $A$ by a factor between 2 and 4.

\subsection{Polarization dependence}
For linear polarization $\hat\bE=(\cos\theta,\sin\theta)$, with $\theta$ measured from the antinodal axis and
$\hat{\mathbf e}_\perp=(-\sin\theta,\cos\theta)$, the tetragonal symmetry of the response $j_i=\chi_{ijkl}E_jE_kE_l$ gives for
each contribution and for the total
\beg\label{PolLaw}
\begin{aligned}
&j_\parallel=A+B\cos4\theta,\qquad j_\perp=-B\sin4\theta,\\
&A=\tfrac34(a+b),\qquad B=\tfrac14(a-3b),
\end{aligned}
\en
with $a=\chi_{xxxx}$, $b=\chi_{xxyy}$. For an arbitrary polarization Eq.~\eqref{PolLaw} is the statement
\beg\label{PolGeneral}
\mathbf j^{(3\omega)}=(A-B)(\bE\cdot\bE)\bE+2B(E_x^2-E_y^2)\bE',
\en
which reduces to Eq.~\eqref{PolLaw} for $\hat\bE=(\cos\theta,\sin\theta)$. Comparison with
Eqs.~(\ref{jHiggs})--(\ref{jqp3w}) gives
\beg\label{AB3w}
\begin{aligned}
A&=j_0\bigg\{\Lambda_{\rm H}\chi_{\rm SH}F_\Delta+\int d\eps\sum_{ab}c_{ab}\Big[\frac34\Gcal^{(4)}_{ab}\\
&\qquad+\frac12{\cal S}_{ab}[\varrho,\Gcal]-\frac18{\cal S}_{ab}[\psi,\Fcal]\Big]\bigg\},\\
B&=j_0\int d\eps\sum_{ab}c_{ab}\Big[\frac14{\cal K}^{(4)}_{ab}-\frac18{\cal S}_{ab}[\psi,\Fcal]\Big].
\end{aligned}
\en
The amplitude mode and the $\ttau3$ vertex contribute only to $A$, the $\itt$ vertex contributes equally to $A$ and $B$,
and the drive term contributes to $A$ through $\Gcal$ and to $B$ through the fourth harmonic ${\cal K}$. The two coefficients have a direct physical meaning. For an isotropic medium $a=3b$ and $B=0$, so $B$ measures the
anisotropy of the nonlinear response. Here it has two sources: the fourth harmonic ${\cal K}$ of the angle-resolved
function $g_{\bn\eps}$, which is nonzero only because the $d$-wave gap makes the quasiparticle spectrum depend on the
direction on the Fermi surface, and the $B_{1g}$ channel of the $2\omega$ state through the $\itt$ vertex. The Higgs
contribution is isotropic, and this gives a polarization signature that separates it from the quasiparticle contribution.

For circular polarization, $\hat\bE=(1,i)/\sqrt2$, Eq.~\eqref{PolGeneral} gives
\beg
\mathbf j^{(3\omega)}=\sqrt2\,B\,E^3\,(1,-i).
\en
In this case the third harmonic has the opposite helicity, its amplitude is set by $B$ alone, and the amplitude mode does not contribute,
since $\delta\Delta_{2\omega}\propto\bE\cdot\bE=0$. For linear polarization along the antinodal and the nodal direction,
$j_\parallel=A+B$ and $A-B$, respectively. In the range of Fig.~\ref{Fig4}, $|B/A|$ lies between $0.04$ and $0.16$. At low
frequency the third harmonic is larger for the field along the nodes and at higher frequency for the field along the
antinodes, with the crossover at $\omega\approx0.3\,\Delta_{\max}$ for $T=0.2\,T_{c0}$ and $\omega\approx0.4\,\Delta_{\max}$ for
$T=0.6\,T_{c0}$.

\section{Discussion}\label{sec:disc}
As we have argued in this paper, the third harmonic response must vanish in the clean limit in a particle-hole symmetric case. 
It is then instructive to ask what the third harmonic measures in the clean limit. For the Galilean invariant spectrum of model \eqref{Eq1} it vanishes identically because with a parabolic dispersion and a pairing interaction that depends only on the relative momenta, a spatially uniform $\mathbf A(t)$ couples only to the total momentum, which is conserved in a clean superconductor. Hence $\mathbf j(t)=-(ne^2/mc)\mathbf A(t)$, with $n$ the electron density, both within and beyond the quasiclassical approximation \cite{Silaev2019,Manske-dwave2020}. A clean superconductor generates a third harmonic only when the band is not parabolic. Expanding $\xi(\bk-e\mathbf A/c)$ to second order in $\mathbf A$, the field then couples through the inverse effective mass projected on the polarization, $D_\bk=\hat\bE\cdot(\partial^2\xi_\bk/\partial\bk\,\partial\bk)\cdot\hat\bE$, which acts as a momentum-dependent scalar potential oscillating at $2\omega$. The average of $D_\bk$ over the Fermi surface is a uniform potential and drops out by gauge invariance once the phase mode is included. The variation of $D_\bk$ around the Fermi surface drives the charge-density fluctuations, and since the density--density bubble is even in $\xi$, this contribution survives exact particle--hole symmetry. The amplitude mode, on the other hand, couples to a density-like perturbation only through a bubble that is odd in $\xi$. The Higgs part of the clean-limit response therefore requires the energy dependence of $D_\bk$ or of the density of states, i.e. particle--hole asymmetry \cite{VladIFE2026}. This is the reason given in Ref.~\cite{Cea2016} for the smallness of the Higgs contribution, and it is consistent with the factor of more than $10^2$ between the two contributions found in Ref.~\cite{Manske-dwave2020}. The anisotropy of $D_\bk$ is tied to the slope of the density of states only for an isotropic two-dimensional band $\xi(k)$. There the anisotropic part of the vertex, $\frac12(\xi''-v_F/p_F)\cos2(\phi_\bk-\theta)$, is fixed by the slope of the density of states per spin, $d\nu/d\xi=(p_F/2\pi v_F^3)(v_F/p_F-\xi'')$, where $v_F=\xi'$ and the derivatives are taken at $k=p_F$. For a lattice dispersion the two are independent, since the vertex contains lattice harmonics, such as $\cos k_xa-\cos k_ya$ for nearest-neighbour hopping with lattice constant $a$, that are not fixed by the slope of the density of states. In an ultra-clean superconductor the third harmonic thus measures how the band deviates from a parabola at the Fermi surface, and only its Higgs part measures the particle--hole asymmetry. For weak disorder in an $s$-wave superconductor the ratio of this response to the impurity-assisted one is about $10^3\,\tau T_c\,(T_c/\veps_F)^2$, with $\tau=1/\Gel$, so that it dominates for $\tau T_c>10^{-3}(\veps_F/T_c)^2$ \cite{Silaev2019}. Although a $d$-wave counterpart of this criterion is not derived here, but we believe it remains quantitatively similar to the $s$-wave case.

The magnitude of the third harmonic in a $d$-wave superconductor is set by momentum relaxation. Both contributions vanish in the clean limit, and at weak disorder the quasiparticle current is linear in $\Gel$ while the Higgs current is quadratic, since the latter requires momentum relaxation twice. Near the gap scale the current is of order $c\,(\Gel/T_{c0})(e\nu_Fv_FT_{c0}/4)(ev_F|\bE|/\omega T_{c0})^3$ with $c$ between $0.1$ and $0.6$. It stays close to linear in $\Gel$ up to $\Gel=0.4\,T_{c0}$, falls off monotonically with frequency between $\omega\approx0.5\,\Delta_{\max}$ and $1.5\,\Delta_{\max}$, and shows no feature at $2\omega=2\Delta_{\max}$. Without the disorder vertex correction the quasiparticle current would be about three times smaller.  For an $s$-wave superconductor at weak disorder, low temperatures and low frequencies Silaev found the same structure \cite{Silaev2019}. Both contributions vanish without impurities, the source of the amplitude oscillation is proportional to $1/\tau$, with $\tau=1/\Gel$, and the quasiparticle and Higgs currents are proportional to $1/\tau$ and $1/\tau^2$, so that the quasiparticle current dominates. The two cases differ in the pair susceptibility. In the $s$-wave case $\chi_{\rm SH}$ is singular at $2\omega=2\Delta$, and in the dirty limit both currents peak at $\omega=\Delta$, with a Higgs peak several times larger than the quasiparticle one \cite{Silaev2019}. In the $d$-wave case the amplitude oscillation decays into nodal quasiparticles at every frequency, $\chi_{\rm SH}$ stays finite at $2\omega=2\Delta_{\max}$, and nothing compensates the additional power of $\Gel$, so that the bound \eqref{jH2jqp} holds at all frequencies of Fig.~\ref{Fig4}. Moreover, the dirty limit is not available for $d$-wave pairing. Nonmagnetic disorder leaves $T_c$ of an $s$-wave superconductor unchanged, so that $1/\tau\gg\Delta$ can be reached, whereas Born scattering is pair breaking for $d$-wave pairing and $T_c$ vanishes at $\Gel\approx1.764\,T_{c0}$, Fig.~\ref{Fig2}. The ratio $\Gel/\Delta$, which sets the relative size of the Higgs contribution, can therefore become large only close to this critical value, where $\Delta$ itself goes to zero.

Two calculations that keep the full lattice dispersion can be compared with ours. Schwarz and Manske \cite{Manske-dwave2020} treated the clean limit, in which the response comes entirely from the band curvature that the quasiclassical theory does not contain. This calculation therefore overlaps with ours only in the polarization dependence. Both give a Higgs contribution that does not depend on the polarization and a transverse current proportional to $\sin4\theta$ that comes from the charge (quasiparticle) channel alone, which is the form required by the tetragonal symmetry. The strength of the anisotropy differs. Their charge-density-fluctuation current is about twice as large for the antinodal as for the nodal polarization, whereas in the range of Fig.~\ref{Fig4} the ratio of our two currents lies between $0.86$ and $1.38$, and the nodal current is the larger one below $\omega\approx0.3\,\Delta_{\max}$. 

Lastly, Benfatto, Castellani and Seibold \cite{Benfatto2023} computed the quasiparticle response of a disordered $d$-wave superconductor on a lattice at $T=0$, with the disorder included explicitly and with both the paramagnetic and the diamagnetic (band-curvature) terms. They found that the paramagnetic, disorder-activated term dominates already at relatively weak disorder, that the response shows no enhancement when the frequency matches the gap maximum, and that the paramagnetic term depends only weakly on the polarization. The impurity-assisted current obtained here is likewise activated by disorder, has no feature at $2\omega=2\Delta_{\max}$, and depends only weakly on the polarization, with $|B/A|$ between $0.04$ and $0.16$ in the range of Fig.~\ref{Fig4}. Their calculation does not include the collective modes, so the ratio \eqref{jH2jqp} of the Higgs and quasiparticle currents has no counterpart there. The band-curvature contribution, which is present also in the normal state, lies outside the quasiclassical theory and is not calculated here.

\section{Conclusions}\label{sec:concl}
We have formulated a Keldysh--Nambu quasiclassical theory of the nonlinear electromagnetic response of a disordered $d$-wave superconductor and applied it to a problem of the third-harmonic generation. For a spatially uniform drive the vector potential can be removed from the collisionless Eilenberger equation by a scalar phase transformation, so that the nonlinear response is built from finite differences, in energy, of the equilibrium propagator. As a consequence the third-harmonic current of a clean superconductor vanishes identically in the collisionless limit, in both the Schmid--Higgs and the charge-fluctuation channels and to all orders in $\Delta$, the spectral and nonequilibrium-distribution contributions cancelling exactly. As a result, the third-harmonic response is determined entirely by momentum relaxation. 

We have further shown that the polarization dependence of the third harmonic is fixed by symmetry independently of the magnitude of either channel: the amplitude-mode contribution is isotropic and purely longitudinal, whereas the particle--hole contribution carries a fourfold modulation and a transverse component $\propto\sin4\theta$, so that any transverse third-harmonic current, and any polarization dependence, is a fingerprint of the particle--hole channel. This angular structure agrees in form with that found by Schwarz and Manske \cite{Manske-dwave2020} for a lattice dispersion, the two calculations addressing complementary --- particle--hole-symmetric and particle--hole-asymmetric --- parts of the response.

A quantitative treatment of the third harmonic with the relaxation self-energy retained consistently in all orders of the hierarchy, the inclusion of the long-range Coulomb interaction, the restoration of band curvature in the current operator, and the experimental signatures of the predicted polarization dependence remain natural directions for future work.

\begin{acknowledgments}
We would like to thank A. F. Volkov for useful comments. This work was financially supported by the National Science Foundation Grant No. DMR-2400484. One of us (MD) has performed the main part of this work at Aspen Center for Physics, which is supported by the National Science Foundation Grant No. PHY-2210452. MD also thanks Niels Bohr Institute where part of this work has been completed for hospitality. The authors acknowledge the use of Anthropic's
Claude for cross-checking the analytical derivations, for independent
numerical verification of the results, and for figure preparation and
manuscript checking. All outputs of Claude were verified and validated by
the authors.

\end{acknowledgments}

\begin{widetext}
\begin{appendix}
\section{Gauge structure of the clean response at $\bq=0$}\label{AppendixA:gauge}
In this Appendix we show that for a spatially uniform vector potential the perturbative solution of the collisionless Eilenberger equation has a simple closed form, which underlies the cancellations in the response functions in the clean superconductor. This demonstrates that for $\bq=0$ disorder provides necessary mechanism for the momentum relaxation leading to the absorption of the electromagnetic radiation by a superconductor. 

Consider the Eilenberger equation \eqref{EilenMain} in the two-time representation, $\check g=\check g(\bn;t_1,t_2)$, before the Wigner transformation to $(\eps,t)$ is performed. The terms $[\eps\check\tau_3\stackrel{\circ},\check g]+\frac{i}{2}\{\check\tau_3,\partial_t\check g\}$ combine into $i\check\tau_3\partial_{t_1}\check g+i\partial_{t_2}\check g\,\check\tau_3$, and the field term is $a_\bn(t_1)\check\tau_3\check g-\check g\check\tau_3a_\bn(t_2)$ with $a_\bn(t)=(ev_F/c)\,\bn\mathbf A(t)$. The equation therefore reads
\beg\label{GaugeEq}
\check\tau_3\big[i\partial_{t_1}+a_\bn(t_1)\big]\check g+\big[i\partial_{t_2}-a_\bn(t_2)\big]\check g\,\check\tau_3+\big[\check\Delta_\bn\stackrel{\circ},\check g\big]=0.
\en
Let $\chi_\bn(t)$ be a scalar function with $\dot\chi_\bn(t)=a_\bn(t)$ and define $\check g(\bn;t_1,t_2)=e^{i\chi_\bn(t_1)}\,\tilde{\check g}(\bn;t_1,t_2)\,e^{-i\chi_\bn(t_2)}$. The scalar phase commutes with $\check\tau_3$ and with $\check\Delta_\bn$, and
$[i\partial_{t_1}+a_\bn(t_1)]e^{i\chi_\bn(t_1)}\tilde{\check g}=e^{i\chi_\bn(t_1)}\,i\partial_{t_1}\tilde{\check g}$, $[i\partial_{t_2}-a_\bn(t_2)]\tilde{\check g}e^{-i\chi_\bn(t_2)}=(i\partial_{t_2}\tilde{\check g})e^{-i\chi_\bn(t_2)}$. Hence $\tilde{\check g}$ satisfies the Eilenberger equation \emph{without} the field, with the same $\check\Delta_\bn$, and the normalization $\check g\circ\check g=\check1$ is preserved by the transformation. If the field is switched on adiabatically and $\check\Delta_\bn$ is held at its equilibrium value, $\tilde{\check g}$ is the equilibrium propagator, $\tilde{\check g}(\bn;t_1,t_2)=\check g_{\bn}(t_1-t_2)$: a uniform vector potential in the collisionless quasiclassical theory is a pure gauge along each trajectory, and the only effect of the field is the time-dependent phase $e^{i[\chi_\bn(t_1)-\chi_\bn(t_2)]}$. The energy shift $a_\bn(t)$ is the Doppler shift of the quasiparticle energy in the moving condensate; the statement is that in the absence of momentum relaxation this shift has no observable consequence for a linearized spectrum at $\bq=0$. The quasiclassical current \eqref{MainCurrent} is $\propto\int d\eps\,\textrm{Tr}[\hat\tau_3\hat g^K(\bn\eps;t)]$ and depends only on the equal-time propagator, for which the phase factors cancel: the current is the equilibrium one, i.e.\ zero, to all orders in the field, apart from the linear (London) term generated by the large-$|\eps|$ behavior $\hat g^K\to2\hat\tau_3\mathrm{sgn}(\eps)$, which the shift of the integration variable does not remove.

The perturbative content of this statement follows by expanding the phase factors. For the monochromatic field \eqref{VectorPotential} we have $a_\bn(t)=\alpha_\bn e^{-i\omega t}+\textrm{c.c.}$ and $\chi_\bn(t)=(i\alpha_\bn/\omega)e^{-i\omega t}+\textrm{c.c.}$ with $\alpha_\bn\equiv(ev_F/i\omega)(\bn\mathbf E)$ (which is ${\cal A}_{+}$ in the main text). In the Wigner representation, $\check g(\bn\eps;t)=\int d\tau\,e^{i\eps\tau}\check g(\bn;t+\tau/2,t-\tau/2)$, a factor $e^{-im\omega t_1}=e^{-im\omega t}e^{-im\omega\tau/2}$ shifts the energy argument of $\check g_\bn(\tau)$ by $-m\omega/2$, while $e^{-im\omega t_2}$ shifts it by $+m\omega/2$. Expanding $e^{i[\chi_\bn(t_1)-\chi_\bn(t_2)]}$ one finds, for the harmonics $e^{-im\omega t}$ with $m=1,2,3$,
\beg\label{FiniteDifferences}
\begin{aligned}
\check g_{1,+}(\bn\eps;\omega)&=\frac{\alpha_\bn}{\omega}\left[\check g_{\bn\eps+\frac{\omega}{2}}-\check g_{\bn\eps-\frac{\omega}{2}}\right],\quad \check g_{2,+}(\bn\eps;\omega)=\frac{\alpha_\bn^2}{2\omega^2}\left[\check g_{\bn\eps+\omega}-2\check g_{\bn\eps}+\check g_{\bn\eps-\omega}\right],\\
\check g_{3,+}(\bn\eps;\omega)&=\frac{\alpha_\bn^3}{6\omega^3}\left[\check g_{\bn\eps+\frac{3\omega}{2}}-3\check g_{\bn\eps+\frac{\omega}{2}}+3\check g_{\bn\eps-\frac{\omega}{2}}-\check g_{\bn\eps-\frac{3\omega}{2}}\right],
\end{aligned}
\en
and for the static second-order part $\check g_{2,0}(\bn\eps)=(|\alpha_\bn|^2/\omega^2)[\check g_{\bn\eps+\omega}-2\check g_{\bn\eps}+\check g_{\bn\eps-\omega}]$, i.e.\ the $m$-th harmonic at $m$-th order is the $m$-th finite difference of the equilibrium propagator with step $\omega$. All components --- retarded, advanced and Keldysh --- obey \eqref{FiniteDifferences}, since the transformation acts on the full Keldysh matrix. 

When the pairing field is allowed to respond, $\check\Delta_\bn\to\check\Delta_\bn+\delta\check\Delta_\bn(t)$, the transformation still removes the vector potential, and $\tilde{\check g}$ is the response of the equilibrium superconductor to $\delta\check\Delta_\bn(t)$ alone. The finite-frequency responses are determined uniquely by \eqref{FiniteDifferences}, whereas the static Keldysh function is not: any zero mode $(\hat g^R_{\bn\eps}-\hat g^A_{\bn\eps})\varphi(\eps)$ can be added to $\hat g_{2,0}^K$ without violating the collisionless equations, and which $\varphi$ is realized depends on the relaxation mechanism. Momentum relaxation breaks the phase invariance. Elastic impurity scattering breaks it through the angular average, which mixes trajectories with different $\chi_\bn$, and band curvature through the $\bk$-dependence of the velocity, which lies outside the quasiclassical equation \eqref{EilenMain}. The clean collisionless response is therefore the point of departure for a controlled expansion in the momentum-relaxation rate rather than a result in itself.

\section{First order corrections}\label{AppendixB:first_order}
In this Section we will compute the first order correction to the quasiclassical propagator \eqref{gexpand}. For definiteness we consider the Fourier component corresponding to $e^{-i\omega t}$ harmonic. From \eqref{EilenMain} expanded to the first order in vector potential at $\bq=0$ we find
\beg\label{A-Eq4g1}
\big[\eps\ttau3+\hat\Delta_\bn,\gh_{1,+}\big]+\frac\omega2\big\{\ttau3,\gh_{1,+}\big\}
-\big[\check{\sigma}_{\rm imp},\gh_{1,+}\big]_\circ
={\cal A}_+\Big[\gh_{\bn\eps+\frac\omega2}\ttau3-\ttau3\gh_{\bn\eps-\frac\omega2}\Big],
\en
with ${\cal A}_+=c{\mathbf E}/(i\omega)$ of \eref{VectorPotential}. By the shift rule \eqref{GM_Shift} the left-hand side equals
$\tveps_{\eps+\frac\omega2}\ttau3\gh_{1,+}-\gh_{1,+}\ttau3\tveps_{\eps-\frac\omega2}+[\hat\Delta_\bn,\gh_{1,+}]$,
i.e.\ both self-energies are absorbed into $\tveps$ of \eref{xitilde} at the two
shifted energies. Using $\tveps_a\ttau3+\hat\Delta_\bn=\eta_a\gh_a$ this is
$\eta_{a}\gh_a\gh_{1,+}-\eta_b\gh_{1,+}\gh_b$ with $a=\eps+\frac\omega2$,
$b=\eps-\frac\omega2$, and together with the first-order normalization
$\gh_a\gh_{1,+}+\gh_{1,+}\gh_b=0$ it becomes
$(\eta_a+\eta_b)\gh_a\gh_{1,+}$. Multiplying by $\gh_a$ gives we find
\beg\label{g1RA}
\gh^{R(A)}_{1,\pm}={\cal A}_\pm\,
\frac{\ttau3-\gh^{R(A)}_{\bn\eps\pm\frac\omega2}\ttau3\,\gh^{R(A)}_{\bn\eps\mp\frac\omega2}}
{\eta^{R(A)}_{\bn\eps+\frac\omega2}+\eta^{R(A)}_{\bn\eps-\frac\omega2}},
\en
Here the $e^{i\omega t}$ harmonic follows from $\omega\to-\omega$, $\bE\to\bE^*$. For the Keldysh component we write $\gh^K_{1,\pm}=\gh^R_{1,\pm}\,t_{\eps\mp\frac\omega2}-\gh^A_{1,\pm}\,t_{\eps\pm\frac\omega2}
+\delta\gh^K_{1,\pm}$ with
\beg\label{g1K}
\begin{aligned}
&\delta\gh^K_{1,\pm}=
\frac{{\cal A}_\pm\big[\ttau3-\gh^{R}_{\bn\eps\pm\frac\omega2}\ttau3\,\gh^{A}_{\bn\eps\mp\frac\omega2}\big]
\big(t_{\eps\pm\frac\omega2}-t_{\eps\mp\frac\omega2}\big)}
{\eta^{R}_{\bn\eps\pm\frac\omega2}+\eta^{A}_{\bn\eps\mp\frac\omega2}} .
\end{aligned}
\en
One inserts the ansatz \eref{g1K} into the Keldysh block of \eref{A-Eq4g1}. As a result, the
terms proportional to $t_{\eps\pm\omega/2}$ cancel against the $R$ and $A$
equations once the $\hs^K$ of the impurities are included, and what remains is
solved as before with $\eta^R_a+\eta^A_b$ in the denominator, giving
$\delta\gh^K_{1,\pm}$ of \eref{g1K}. 

These are the clean-limit expressions with $\eta_{\bn\eps}^{R(A)}$ are the functions of  $\tveps$ \eqref{xitilde}. The
reason is that the equilibrium part of the elastic self-energy
$-{i\gamma_{\rm el}}\Gcal_\eps\ttaut/2$ is time independent and proportional to
$\ttaut$, so that in the first-order equation its $\circ$-product with $\gh_1$
reduces to ordinary products with the
energy argument shifted by $\pm\omega/2$ exactly as for the term $\eps\ttaut$ alone. The two terms $\eps\ttaut$ and $({i\gamma_{\rm el}}/{2})\Gcal_\eps\ttaut$ then combine into
$\tveps_{\eps\pm\omega/2}\ttaut$ and no separate impurity term survives. A
first-order correction to the self-energy itself, $\propto\avg{\gh_1}$, is
absent because $\gh_{1,\pm}\propto(\bn\cdot\bE)$ is odd under $\bn\to-\bn$.

Expressions above can be re-written using the identity
\beg\label{SimpleIdentity}
\gh_a\ttau3\gh_b=(g_ag_b+f_af_b)\ttau3+(g_af_b+f_ag_b)i\hat{\tau}_2, 
\en
which uses
$\ttau3\,\itt=\ttau1=-\itt\,\ttau3$ and $(\itt)^2=-1$. Then it obtains
\begin{align}
\gh_{1,\pm}^{R(A)}(\bn\eps)&=-{\cal A}_\pm\frac{B^{R(A)}_{\bn}(\eps+\frac\omega2,\eps-\frac\omega2)\ttau3
+C^{R(A)}_{\bn}(\eps+\frac\omega2,\eps-\frac\omega2)\itt}{\eta^{R(A)}_{\bn\eps+\frac\omega2}+\eta^{R(A)}_{\bn\eps-\frac\omega2}},\label{A-blocks}\\
\delta\gh^K_{1,\pm}(\bn\eps)&=-{\cal A}_\pm\frac{\big[B^{RA}_\bn(\eps\pm\frac\omega2,\eps\mp\frac\omega2)\ttau3
+C^{RA}_\bn(\eps\pm\frac\omega2,\eps\mp\frac\omega2)\itt\big]\big(t_{\eps\pm\frac\omega2}-t_{\eps\mp\frac\omega2}\big)}
{\eta^{R}_{\bn\eps\pm\frac\omega2}+\eta^{A}_{\bn\eps\mp\frac\omega2}},\label{A-blocksK}
\end{align}
with the notations
\beg\label{BCdef}
\begin{aligned}
&B_{\bn}^{ab}(\eps,\eps')\equiv g_{\bn\eps}^ag_{\bn\eps'}^b+f_{\bn\eps}^af_{\bn\eps'}^b-1,\quad \overline{B}_{\bn}^{ab}(\eps,\eps')\equiv g_{\bn\eps}^ag_{\bn\eps'}^b+f_{\bn\eps}^af_{\bn\eps'}^b+1, \quad C_{\bn}^{ab}(\eps,\eps')\equiv g_{\bn\eps}^af_{\bn\eps'}^b+f_{\bn\eps}^ag_{\bn\eps'}^b.
\end{aligned}
\en
We will use these expressions to compute the second and third order corrections.

\section{Second order corrections}\label{AppendixC}
In this Appendix we derive the results of Sec.~\ref{sec:2w}. We write $\gh^a_s\equiv\gh^a_{\bn,\eps+s\omega}$,
$\eta^a_s$, $\tveps^a_s$ and $t_s\equiv t_{\eps+s\omega}$ with $s=+,0,-$, and $H^a_s\equiv\eta^a_s\gh^a_s=\tveps^a_s\ttau3+\hat\Delta_\bn$.

\paragraph{Keldysh rotation.} Every equilibrium matrix $\check A_\eps$ with $\hat A^K_\eps=(\hat A^R_\eps-\hat A^A_\eps)t_\eps$,
that is $\gc_{\bn\eps}$, $\avg{\gc_{\bn\eps}}$, $\check\sigma_{\rm imp}(\eps)$ and $\check H_\eps$, is brought to block-diagonal
form by
\beg\label{KeldyshRotation}
\check L_\eps=\begin{pmatrix}1&-t_\eps\\0&1\end{pmatrix},\qquad
\check L_\eps^{-1}\check A_\eps\check L_\eps={\rm diag}\big(\hat A^R_\eps,\hat A^A_\eps\big).
\en
A function that connects the energies $\eps_1$ (left) and $\eps_2$ (right), such as $\gc_2$ with $\eps_{1,2}=\eps\pm\omega$
or $\gc_1(\bn,\eps')$ with $\eps_{1,2}=\eps'\pm\omega/2$, transforms as $\check L^{-1}_{\eps_1}\check O\check L_{\eps_2}$, and
products of such functions transform factor by factor. For $\gc_2$ this gives the diagonal blocks $\gh_2^R$, $\gh_2^A$ and
the off-diagonal block $\delta\gh_2^K$ of Eq.~\eqref{g2Kansatz2w}. The first-order functions acquire the diagonal blocks
$\gh^{R}_1$, $\gh^{A}_1$ and the off-diagonal block $\delta\gh^K_1$ of Eq.~\eqref{g1K}. The matrix $\check\tau_3$ between
the energies $\eps_1$ and $\eps_2$ acquires the off-diagonal block $(t_{\eps_1}-t_{\eps_2})\ttau3$, and $\delta\check\Delta_\bn$
the off-diagonal block $w_{RA}\dDh_\bn$. The Born vertex is linear in $\gc_2$, so that its blocks are
$\delta\hat\sigma_{ab}=-(i\Gel/2)\avg{X_{ab}}$. Equations (\ref{Eq2w},\ref{Norm2w}) then take the form \eqref{SectorEq}
with
\beg\label{Sources2w}
\begin{aligned}
\hat S_{RR}&={\cal A}_+\big[\gh^R_1(\eps+\tfrac\omega2)\ttau3-\ttau3\gh^R_1(\eps-\tfrac\omega2)\big],\qquad
\hat Z_{RR}=\gh^R_1(\eps+\tfrac\omega2)\,\gh^R_1(\eps-\tfrac\omega2),\\
\hat S_{RA}&={\cal A}_+\big[\gh^R_1(\eps+\tfrac\omega2)\ttau3\,(t_0-t_-)+\ttau3\gh^A_1(\eps-\tfrac\omega2)\,(t_0-t_+)
+\delta\gh^K_1(\eps+\tfrac\omega2)\ttau3-\ttau3\,\delta\gh^K_1(\eps-\tfrac\omega2)\big],\\
\hat Z_{RA}&=\gh^R_1(\eps+\tfrac\omega2)\,\delta\gh^K_1(\eps-\tfrac\omega2)+\delta\gh^K_1(\eps+\tfrac\omega2)\,\gh^A_1(\eps-\tfrac\omega2),
\end{aligned}
\en
and $\hat S_{AA}$, $\hat Z_{AA}$ obtained from $\hat S_{RR}$, $\hat Z_{RR}$ by $R\to A$. Here
$\gh^{R(A)}_1(\eps')\equiv\gh^{R(A)}_{1,+}(\bn\eps';\omega)$ and $\delta\gh^K_1(\eps')\equiv\delta\gh^K_{1,+}(\bn\eps';\omega)$ are
given by Eqs.~(\ref{g1RA},\ref{g1K}).

\paragraph{General expressions.} Equations (\ref{Eq2w},\ref{Norm2w}) then split into three problems of the same form for the matrices
$X_{RR}=\gh_2^R$, $X_{AA}=\gh_2^A$ and $X_{RA}=\delta\gh_2^K$. With the short-hand
notation $\gh^a_\pm\equiv\gh^a_{\bn,\eps\pm\omega}$ and $\eta^a_\pm\equiv\eta^a_{\bn,\eps\pm\omega}$ for $a=R,A$ they read
\beg\label{SectorEq}
\begin{aligned}
&\eta^a_+\gh^a_+X_{ab}-X_{ab}\,\eta^b_-\gh^b_-=\hat S_{ab}
-w_{ab}\big(\dDh_\bn\gh^b_--\gh^a_+\dDh_\bn\big)
+\delta\hat\sigma_{ab}\gh^b_--\gh^a_+\delta\hat\sigma_{ab},\\
&\gh^a_+X_{ab}+X_{ab}\,\gh^b_-=-\hat Z_{ab},
\end{aligned}
\en
where $\dDh_\bn=\itt\,\Ycal_\bn\delta\Delta_{2\omega}$, $w_{RR}=w_{AA}=1$, $w_{RA}=t_{\eps+\omega}-t_{\eps-\omega}$,
$\delta\hat\sigma_{ab}=-(i\Gel/2)\avg{X_{ab}}$, $ab=RR,AA,RA$ and we also used $\eta^a\gh^a=\tveps^a\ttau3+\hat\Delta_\bn$, which
follows directly from Eq.~\eqref{gnRA}. 
The drive terms $\hat S_{ab}$ and $\hat Z_{ab}$ are built from the
first-order functions and are given in  Eq.~\eqref{Sources2w}. The weight $w_{RA}$ is the difference of the equilibrium distribution functions $t_\eps$ at the two
energies $\eps\pm\omega$ connected by the oscillating pairing field. It vanishes for a static perturbation.

The solution of Eqs.~\eqref{SectorEq} is most compactly written in terms of divided differences. They are built from the
equilibrium functions at the energies $\eps+s\omega$ with $s=+,0,-$, taken on the retarded ($a=R$) or the advanced ($a=A$)
branch, and from the corresponding complex energies $\tveps^a_s\equiv\tveps^a_{\eps+s\omega}$ of Eq.~\eqref{xitilde} in the main text. In order to keep the resulting expressions compact, we use
the label $a_s$ for the branch $a$ at the energy $\eps+s\omega$, so that, for example, $R_+$ refers to the retarded branch
at $\eps+\omega$ and $A_0$ to the advanced branch at $\eps$. For an equilibrium quantity $Q^a_\eps$, which can be a matrix
such as $\gh^a_{\bn\eps}$ or a number such as $\Gcal^a_\eps$, the first and second divided differences are defined as
\beg\label{DivDiff}
\begin{aligned}
Q[a_s,b_{s'}]&=\frac{Q^a_{\eps+s\omega}-Q^b_{\eps+s'\omega}}{\tveps^a_s-\tveps^b_{s'}},\quad
Q[a_s,b_{s'},c_{s''}]=\frac{Q[a_s,b_{s'}]-Q[b_{s'},c_{s''}]}{\tveps^a_s-\tveps^c_{s''}},
\end{aligned}
\en
where each of $a$, $b$, $c$ is $R$ or $A$, and each of $s$, $s'$, $s''$ is $+$, $0$ or $-$. For example,
$\gh[R_+,A_0]=(\gh^R_{\bn,\eps+\omega}-\gh^A_{\bn\eps})/(\tveps^R_+-\tveps^A_0)$. In the clean limit $\tveps^a_s\to\eps+s\omega$,
and the divided differences become finite differences with step $\omega$.

Without the disorder vertex correction and for $\delta\Delta_{2\omega}=0$ the solution of Eqs.~\eqref{SectorEq} can be easily obtained:
\beg\label{DriveDD}
\begin{aligned}
X^{(0)}_{RR}&={\cal A}_+^2\,\gh[R_+,R_0,R_-],\quad X^{(0)}_{AA}={\cal A}_+^2\,\gh[A_+,A_0,A_-],\\
X^{(0)}_{RA}&={\cal A}_+^2\big\{(t_{\eps+\omega}-t_\eps)\,\gh[R_+,A_0,A_-]
+(t_\eps-t_{\eps-\omega})\,\gh[R_+,R_0,A_-]\big\}.
\end{aligned}
\en
In the clean limit Eq.~\eqref{DriveDD} reduces to the second finite difference
of Eq.~\eqref{FiniteDifferences}, which is a consequence of the gauge structure of the collisionless problem
(Appendix~\ref{AppendixA:gauge}). Elastic scattering enters the drive part only through the positions $\tveps^a_s$ of the
points in the complex energy plane.

The pairing field and the disorder vertex add the responses to commutator-type sources (Appendix~\ref{AppendixC}).
For $d$-wave pairing $\avg{X_{ab}}=X^{(3)}_{ab}\ttau3+X^{(2)}_{ab}\itt$, and the full solution reads
\beg\label{Solution2w}
\begin{aligned}
X_{ab}&=X^{(0)}_{ab}+\frac{i\Gel}{2}\,X^{(3)}_{ab}\,\gh[a_+,b_-]+\Big(w_{ab}\Ycal_\bn\,\delta\Delta_{2\omega}+\frac{i\Gel}{2}X^{(2)}_{ab}\Big)\frac{\bar B_{ab}\,\itt+C_{ab}\,\ttau3}{\eta^a_++\eta^b_-},
\end{aligned}
\en
with $\bar B_{ab}=g^a_+g^b_-+f^a_+f^b_-+1$ and $C_{ab}=g^a_+f^b_-+f^a_+g^b_-$, which are fairly standard combinations in quasiclassical theories \cite{MoorVolkovEfetov2017,LiDzero2024,Eremin2024}. The second term is the
response to the isotropic $\ttau3$ part of $\delta\hat\sigma_{ab}$, which acts as an induced scalar potential. The third
term is the response to the pairing field, which consists of the $d$-wave amplitude $\delta\Delta_{2\omega}$ and of the
isotropic $\itt$ part of $\delta\hat\sigma_{ab}$, an induced $s$-wave pair potential.

The numbers $X^{(3)}_{ab}$ and $X^{(2)}_{ab}$ follow from the Fermi-surface average of Eq.~\eqref{Solution2w}. The drive
enters through ${\cal A}_+^2=-(ev_F/\omega)^2(\bn\cdot\bE)^2$. For any function $h$ of $\Ycal_\bn^2$
\beg\label{PolAverages}
\begin{aligned}
\avg{(\bn\cdot\bE)^2h}&=\left(\frac{E_x^2+E_y^2}{2}\right)\avg{h},\quad
\avg{(\bn\cdot\bE)^2\,\Ycal_\bn h}=\left(\frac{E_x^2-E_y^2}{2\sqrt2}\right)\avg{\Ycal_\bn^2h},
\end{aligned}
\en
where $E_{x,y}$ are the components of $\bE$ along the antinodal axes. The averages then involve, besides
$\Gcal^a_\eps=\avg{g^a_{\bn\eps}}$ of Eq.~\eqref{Avbn}, the function
\beg\label{FcalDef}
\Fcal^a_\eps=\avg{\Ycal_\bn f^a_{\bn\eps}}
\en
that appears in the gap equation, and we find
\beg\label{WardClosure}
\begin{aligned}
X^{(3)}_{ab}=\frac{1}{\kappa_{ab}}\bigg[&-\Big(\frac{ev_F}{\omega}\Big)^{\!2}\left(\frac{E_x^2+E_y^2}{2}\right)\,\Gcal^{(3)}_{ab}
-w_{ab}\,\Fcal[a_+,b_-]\,\delta\Delta_{2\omega}\bigg],
\end{aligned}
\en
\beg\label{ClosureX2}
\begin{aligned}
X^{(2)}_{ab}&=-\Big(\frac{ev_F}{\omega}\Big)^{\!2}\left(\frac{E_x^2-E_y^2}{2\sqrt2}\right)\,\frac{\Fcal^{(3)}_{ab}}{\kappa'_{ab}}.
\end{aligned}
\en
Here $\Gcal^{(3)}_{ab}$ and $\Fcal^{(3)}_{ab}$ are obtained from $X^{(0)}_{ab}/{\cal A}_+^2$ of Eq.~\eqref{DriveDD} by
replacing $\gh$ with $\Gcal$ and $\Fcal$ and 
\beg\label{kappaab}
\kappa'_{ab}=1-\frac{i\Gel}{2}\avg{\frac{\bar B_{ab}}{\eta^a_++\eta^b_-}}.
\en
The prefactor of
Eq.~\eqref{WardClosure} is the inverse of the vertex denominator $\kappa_{ab}=1-(i\Gel/2)\Gcal[a_+,b_-]$, which can be reduced to $\kappa_{ab}=2\omega/(\tveps^a_+-\tveps^b_-)$ by employing a Ward identity below.

Three properties follow. (i) $\kappa_{ab}$ does not vanish for $\omega\neq0$, so that elastic scattering alone determines the
$2\omega$ response. This is to be expected that static disorder is the only source of momentum relaxation in our model, which means the light can only be absorbed with nonzero $\gamma_{\rm el}$.  For the static component of $\gc_2$ the vertex denominator of the $RA$ sector vanishes, and the static
distribution is fixed only by inelastic relaxation \cite{Aidan2026}. (ii) $X^{(3)}_{ab}$ and $\delta\Delta_{2\omega}$
(see below) are proportional to $\bE\cdot\bE=E_x^2+E_y^2$. They do not depend on the direction of linear polarization and
vanish for circular polarization. $X^{(2)}_{ab}$ is proportional to $E_x^2-E_y^2$, that is to $\cos2\theta$ for linear
polarization at the angle $\theta$ to the $x$ axis, and it is finite for circular polarization. (iii) $X^{(2)}_{ab}$ does
not enter the gap equation, but it enters the third-order equations through $\delta\check\sigma$,
Eq.~\eqref{VertexSource3}.

\paragraph{Solution of a sector equation.} Let $a_1$ and $b_2$ be two points, $H^a_1=\eta^a_1\gh^a_1$ and
$H^b_2=\eta^b_2\gh^b_2$, with $(\gh^a_1)^2=(\gh^b_2)^2=1$, and let $\bar B_{ab}=g^a_1g^b_2+f^a_1f^b_2+1$ and
$C_{ab}=g^a_1f^b_2+f^a_1g^b_2$, which for $a_1=a_+$, $b_2=b_-$ are the functions of Eq.~\eqref{Solution2w}. The system $H^a_1X-XH^b_2=\hat S$, $\gh^a_1X+X\gh^b_2=-\hat Z$ has
the unique solution
\beg\label{ClosedForm2w}
X=\frac{\gh^a_1\big(\hat S-\eta^b_2\hat Z\big)}{\eta^a_1+\eta^b_2},
\en
which follows by eliminating $XH^b_2=\eta^b_2X\gh^b_2$ with the second equation. A source of the commutator form
$\hat S=\hat Q\gh^b_2-\gh^a_1\hat Q$ with $\hat Z=0$ gives $X=V_{ab}[\hat Q]\equiv(\gh^a_1\hat Q\gh^b_2-\hat Q)/(\eta^a_1+\eta^b_2)$.
Since $H^a_1-H^b_2=(\tveps^a_1-\tveps^b_2)\ttau3$, the matrix $\gh^a_1-\gh^b_2$ solves the system with
$\hat S=(\tveps^a_1-\tveps^b_2)(\gh^a_1\ttau3-\ttau3\gh^b_2)$ and $\hat Z=0$, which gives the Ward identity
\beg\label{WardMatrix}
V_{ab}[\ttau3]=\frac{\gh^a_1\ttau3\gh^b_2-\ttau3}{\eta^a_1+\eta^b_2}=-\gh[a_1,b_2].
\en
From $\gh^a_1\ttau3\gh^b_2=(g^a_1g^b_2+f^a_1f^b_2)\ttau3+(g^a_1f^b_2+f^a_1g^b_2)\itt$ and
$\gh^a_1\itt\gh^b_2=-(g^a_1g^b_2+f^a_1f^b_2)\itt-(g^a_1f^b_2+f^a_1g^b_2)\ttau3$ it follows that
$(g^a_1g^b_2+f^a_1f^b_2-1)/(\eta^a_1+\eta^b_2)=-g[a_1,b_2]$, $C_{ab}/(\eta^a_1+\eta^b_2)=-f[a_1,b_2]$ and
$V_{ab}[\itt]=-(\bar B_{ab}\itt+C_{ab}\ttau3)/(\eta^a_1+\eta^b_2)$. Applied to Eqs.~(\ref{g1RA},\ref{g1K}), Eq.~\eqref{WardMatrix}
gives the first-order functions in the form
\beg\label{FirstOrderDD}
\gh^{a}_{1,+}(\bn\eps;\omega)={\cal A}_+\,\gh\big[a_{\eps+\frac\omega2},a_{\eps-\frac\omega2}\big]\quad(a=R,A),\qquad
\delta\gh^K_{1,+}(\bn\eps;\omega)={\cal A}_+\big(t_{\eps+\frac\omega2}-t_{\eps-\frac\omega2}\big)\,\gh\big[R_{\eps+\frac\omega2},A_{\eps-\frac\omega2}\big],
\en
where the divided differences are those of Eq.~\eqref{DivDiff} with the points at the energies $\eps\pm\omega/2$. In the
clean limit Eq.~\eqref{FirstOrderDD} reduces to the first finite difference of Eq.~\eqref{FiniteDifferences}.

\paragraph{Drive part.} For any three points $1$, $2$, $3$, each with its own branch, the divided differences of $\gh$ obey
\beg\label{DDidentities}
\begin{aligned}
&H_1\gh[1,2]-\gh[1,2]H_2=\gh_1\ttau3-\ttau3\gh_2,\qquad \gh_1\gh[1,2]+\gh[1,2]\gh_2=0,\\
&H_1\gh[1,2,3]-\gh[1,2,3]H_3=\gh[1,2]\ttau3-\ttau3\gh[2,3],\qquad
\gh_1\gh[1,2,3]+\gh[1,2,3]\gh_3=-\gh[1,2]\,\gh[2,3].
\end{aligned}
\en
The first line follows from $H_1-H_2=(\tveps_1-\tveps_2)\ttau3$ and $\gh_i^2=1$. The second line follows from the first one
with $H_1=H_2+(\tveps_1-\tveps_2)\ttau3$ and $H_3=H_2+(\tveps_3-\tveps_2)\ttau3$. With Eq.~\eqref{FirstOrderDD} the drive terms
\eqref{Sources2w} become
\beg\label{SourcesDD}
\begin{aligned}
&\hat S_{RR}={\cal A}_+^2\big(\gh[R_+,R_0]\ttau3-\ttau3\gh[R_0,R_-]\big),\qquad \hat Z_{RR}={\cal A}_+^2\,\gh[R_+,R_0]\,\gh[R_0,R_-],\\
&\hat S_{RA}={\cal A}_+^2\Big\{(t_+-t_0)\big(\gh[R_+,A_0]\ttau3-\ttau3\gh[A_0,A_-]\big)
+(t_0-t_-)\big(\gh[R_+,R_0]\ttau3-\ttau3\gh[R_0,A_-]\big)\Big\},\\
&\hat Z_{RA}={\cal A}_+^2\Big\{(t_+-t_0)\,\gh[R_+,A_0]\,\gh[A_0,A_-]+(t_0-t_-)\,\gh[R_+,R_0]\,\gh[R_0,A_-]\Big\}.
\end{aligned}
\en
By the second line of Eq.~\eqref{DDidentities}, Eq.~\eqref{DriveDD} solves Eqs.~\eqref{SectorEq} with
$\delta\hat\sigma_{ab}=0$ and $\delta\Delta_{2\omega}=0$, and by the uniqueness of Eq.~\eqref{ClosedForm2w} it is the drive
part of the solution.

\paragraph{Vertex and closure.} The matrices $\avg{X_{ab}}$ have no $\ttau0$ and $\ttau1$ components (see the remark after
Eq.~\eqref{DefineOP}), so that $\delta\hat\sigma_{ab}=-(i\Gel/2)(X^{(3)}_{ab}\ttau3+X^{(2)}_{ab}\itt)$. The solution of
Eqs.~\eqref{SectorEq} is then $X_{ab}=X^{(0)}_{ab}-w_{ab}V_{ab}[\dDh_\bn]+V_{ab}[\delta\hat\sigma_{ab}]$, which is
Eq.~\eqref{Solution2w}. In its Fermi-surface average the terms odd in $\Ycal_\bn$ drop out: $g$, $\eta$, $\bar B_{ab}$ and
$\Ycal_\bn f$ are even in $\Ycal_\bn$, while $f$ and $C_{ab}$ are odd. Equation \eqref{PolAverages} follows from
$(\bn\cdot\bE)^2=\frac12\bE\cdot\bE+\frac12(E_x^2-E_y^2)\cos2\phi_\bn+E_xE_y\sin2\phi_\bn$ and $\Ycal_\bn=\sqrt2\cos2\phi_\bn$,
since $\avg{\sin2\phi_\bn h}=0$ for any function $h$ of $\Ycal_\bn$ ($\phi_\bn\to-\phi_\bn$) and $\avg{\cos2\phi_\bn h}=0$ for any
function $h$ of $\Ycal_\bn^2$ ($\phi_\bn\to\phi_\bn+\pi/2$). With
$\avg{\Ycal_\bn C_{ab}/(\eta^a_++\eta^b_-)}=-\Fcal[a_+,b_-]$ the $\ttau3$ component gives
\beg\label{ClosureGeneral}
\kappa_{ab}X^{(3)}_{ab}=-\Big(\frac{ev_F}{\omega}\Big)^{\!2}\frac{\bE\cdot\bE}{2}\,\Gcal^{(3)}_{ab}-w_{ab}\Fcal[a_+,b_-]\,\delta\Delta_{2\omega},
\qquad \kappa_{ab}=1-\frac{i\Gel}{2}\Gcal[a_+,b_-],
\en
and the $\itt$ component gives Eq.~\eqref{ClosureX2}. In the Born limit $\Phi[\Gcal]=(i\Gel/2)\Gcal$ in Eq.~\eqref{xitilde},
so that $\tveps^a_s=\eps+s\omega+(i\Gel/2)\Gcal^a_s$ for both branches and
\beg\label{WardKappa}
\kappa_{ab}=1-\frac{(\tveps^a_+-\tveps^b_-)-2\omega}{\tveps^a_+-\tveps^b_-}=\frac{2\omega}{\tveps^a_+-\tveps^b_-}.
\en
Together with Eq.~\eqref{ClosureGeneral} this is Eq.~\eqref{WardClosure}. The same steps for the static component, with
equal energies on both sides, give a vanishing numerator. In the $RA$ sector the denominator stays finite,
$\tveps^R_\eps-\tveps^A_\eps=2i\Im\tveps^R_\eps$, so that $\kappa_{RA}=0$ and the closure is singular.

\paragraph{Gap equation.} With $[\hat X]_{\itt}=-\frac12\Tr[\itt\hat X]$ the $\Ycal$-projection of the $\itt$ component of
Eq.~\eqref{Solution2w} is
\beg\label{VertexProjection}
\avg{\Ycal_\bn[X_{ab}]_{\itt}}=-\Big(\frac{ev_F}{\omega}\Big)^{\!2}\frac{\bE\cdot\bE}{2}\,\Fcal^{(3)}_{ab}
+\frac{i\Gel}{2}\Fcal[a_+,b_-]\,X^{(3)}_{ab}+w_{ab}\,\delta\Delta_{2\omega}\avg{\frac{\Ycal_\bn^2\bar B_{ab}}{\eta^a_++\eta^b_-}}.
\en
$X^{(2)}_{ab}$ drops out because its coefficient $\avg{\Ycal_\bn\bar B_{ab}/(\eta^a_++\eta^b_-)}$ vanishes. Since
$\Tr[(-\itt)\gh_2^K]=2\sum_{ab}c_{ab}[X_{ab}]_{\itt}$, Eq.~\eqref{Self4Eli} becomes
$(8/\lam)\,\delta\Delta_{2\omega}=2\int d\eps\sum_{ab}c_{ab}\avg{\Ycal_\bn[X_{ab}]_{\itt}}$, and Eq.~\eqref{WardClosure} for
$X^{(3)}_{ab}$ gives Eqs.~(\ref{Eq4dDelta2w})--(\ref{FDelta2w}). The equilibrium gap equation follows from
Eq.~\eqref{Self4Eli} with $\Tr[(-\itt)\gh^K_{\bn\eps}]=2(f^R_{\bn\eps}-f^A_{\bn\eps})t_\eps$. At large $|\eps|$ its integrand
$(\Fcal^R_\eps-\Fcal^A_\eps)t_\eps/\Delta$ behaves as $2/|\eps|$. This tail is cancelled by the $RR$ and $AA$ terms of
Eq.~\eqref{chiSH2w}, while $c_{RA}w_{RA}=t_{\eps+\omega}-t_{\eps-\omega}$ decays exponentially. The remaining integrand of
$\chi_{\rm SH}^{-1}$ decays as $\eps^{-2}$ and that of $F_\Delta$ as $\eps^{-3}$, so that both integrals converge and
$\omega_D$ can be sent to infinity.

\paragraph{Matrices needed in the third order.} The third-order equations require $\gc_2$ itself. Its blocks are
$\gh_2^R=X_{RR}$, $\gh_2^A=X_{AA}$ and $\gh_2^K=X_{RR}t_{\eps-\omega}-t_{\eps+\omega}X_{AA}+X_{RA}$, with the components
\beg\label{Components2w}
\begin{aligned}
[X_{ab}]_{\ttau3}&={\cal A}_+^2\,g^{(3)}_{ab}+\frac{i\Gel}{2}X^{(3)}_{ab}\,g[a_+,b_-]
+\Big(w_{ab}\Ycal_\bn\delta\Delta_{2\omega}+\frac{i\Gel}{2}X^{(2)}_{ab}\Big)\frac{C_{ab}}{\eta^a_++\eta^b_-},\\
[X_{ab}]_{\itt}&={\cal A}_+^2\,f^{(3)}_{ab}+\frac{i\Gel}{2}X^{(3)}_{ab}\,f[a_+,b_-]
+\Big(w_{ab}\Ycal_\bn\delta\Delta_{2\omega}+\frac{i\Gel}{2}X^{(2)}_{ab}\Big)\frac{\bar B_{ab}}{\eta^a_++\eta^b_-},
\end{aligned}
\en
where $g^{(3)}_{ab}$ and $f^{(3)}_{ab}$ are the combinations of Eq.~\eqref{DriveDD} for $g$ and $f$, and
$[X_{ab}]_{\ttau0}=[X_{ab}]_{\ttau1}=0$. The vertex enters Eq.~\eqref{VertexSource3} through $\delta\check\sigma$ with the blocks
$\delta\hat\sigma^R=\delta\hat\sigma_{RR}$, $\delta\hat\sigma^A=\delta\hat\sigma_{AA}$ and
$\delta\hat\sigma^K=\delta\hat\sigma_{RR}t_{\eps-\omega}-t_{\eps+\omega}\delta\hat\sigma_{AA}+\delta\hat\sigma_{RA}$.


\section{Third order corrections}\label{app:third}
\subsection{Summary of the results}
We keep the notation of Appendix~\ref{AppendixC}, now for
the four energies $\eps+s\omega$ with $s=\pm\frac12,\pm\frac32$: $a_s$ is the branch $a$ at the energy $\eps+s\omega$, and
$t_s\equiv t_{\eps+s\omega}$. The third divided difference is defined by the recursion of Eq.~\eqref{DivDiff},
\beg\label{DivDiff3}
Q[a_s,b_{s'},c_{s''},d_{s'''}]=\frac{Q[a_s,b_{s'},c_{s''}]-Q[b_{s'},c_{s''},d_{s'''}]}{\tveps^a_s-\tveps^d_{s'''}}.
\en

Without the disorder vertex and for $\delta\Delta_{2\omega}=0$ the solution is the third divided difference,
\beg\label{DriveDD3}
\begin{aligned}
\Xi^{(0)}_{RR}&={\cal A}_+^3\,\gh[R_{3/2},R_{1/2},R_{-1/2},R_{-3/2}],\\
\Xi^{(0)}_{AA}&={\cal A}_+^3\,\gh[A_{3/2},A_{1/2},A_{-1/2},A_{-3/2}],\\
\Xi^{(0)}_{RA}&={\cal A}_+^3\big\{(t_{3/2}-t_{1/2})\,\gh[R_{3/2},A_{1/2},A_{-1/2},A_{-3/2}]\\
&+(t_{1/2}-t_{-1/2})\,\gh[R_{3/2},R_{1/2},A_{-1/2},A_{-3/2}]\\
&+(t_{-1/2}-t_{-3/2})\,\gh[R_{3/2},R_{1/2},R_{-1/2},A_{-3/2}]\big\}.
\end{aligned}
\en
The pattern is that of Eq.~\eqref{DriveDD}. Each of the three steps of size $\omega$ that lead from $\eps+3\omega/2$ to
$\eps-3\omega/2$ can carry the transition from the retarded to the advanced branch, with the weight given by the
difference of $t_s$ at the two energies this step connects. In the clean limit Eq.~\eqref{DriveDD3} reduces to the
third finite difference of Eq.~\eqref{FiniteDifferences}.

The pairing field and the disorder vertex enter through the $2\omega$ potential
\beg\label{Potential2w}
\hat U_{ab}(\eps)=\delta\hat\sigma_{ab}-w_{ab}\dDh_\bn,
\en
which appears in Eq.~\eqref{SectorEq} in the commutator form $\hat U_{ab}\gh^b_--\gh^a_+\hat U_{ab}$. For any two points
$1$, $2$ and any matrix $\hat Q$ we write $V[1,2](\hat Q)=(\gh_1\hat Q\gh_2-\hat Q)/(\eta_1+\eta_2)$, which is the
function $V_{ab}[\hat Q]$ of Appendix~\ref{AppendixC}. In this notation Eq.~\eqref{Solution2w} reads
$X_{ab}=X^{(0)}_{ab}+V[a_+,b_-](\hat U_{ab})$, and the Ward identity \eqref{WardMatrix} is $V[1,2](\ttau3)=-\gh[1,2]$.
At third order we also need the divided differences of $V$ with respect to either of its two points,
\beg\label{VDD}
\begin{aligned}
V[1,2;3](\hat Q)&=\frac{V[1,3](\hat Q)-V[2,3](\hat Q)}{\tveps_1-\tveps_2},\quad V[1;2,3](\hat Q)=\frac{V[1,2](\hat Q)-V[1,3](\hat Q)}{\tveps_2-\tveps_3},
\end{aligned}
\en
which for $\hat Q=\ttau3$ both reduce to $-\gh[1,2,3]$. With $\hat U^\pm_{ab}\equiv\hat U_{ab}(\eps\pm\omega/2)$ the
retarded function is
\beg\label{Solution3w}
\begin{aligned}
\Xi_{RR}&=\Xi^{(0)}_{RR}+{\cal A}_+\Big\{V[R_{3/2},R_{1/2};R_{-3/2}]\big(\hat U^-_{RR}\big)+V[R_{3/2};R_{-1/2},R_{-3/2}]\big(\hat U^+_{RR}\big)\Big\},
\end{aligned}
\en
$\Xi_{AA}$ follows by $R\to A$, and $\Xi_{RA}$ is given in Eq.~\eqref{Solution3wRA} below.

The two terms in the curly brackets correspond to the two orders in which the field and the $2\omega$ potential connect
the energies $\eps+3\omega/2$ and $\eps-3\omega/2$. In the first term the field acts first, from $\eps+3\omega/2$ to
$\eps+\omega/2$, and the potential centred at $\eps-\omega/2$ completes the transition. In the second term the potential
centred at $\eps+\omega/2$ acts first. In both cases the field produces the divided difference with respect to the
point at which it acts, which is also the rule that leads from Eq.~\eqref{DriveDD} to Eq.~\eqref{DriveDD3}. In the $RA$
sector each of the two steps can carry the transition from the retarded to the advanced branch. The step of the field
then carries the difference of $t_s$, as in Eq.~\eqref{DriveDD3}, and the step of the potential carries the off-diagonal
block $\hat U_{RA}$, which contains $w_{RA}\dDh_\bn$ and $\delta\hat\sigma_{RA}$.

In the clean limit $\hat U_{RR}$ reduces to $-\dDh_\bn$ at both centre energies, the points move to the real axis, and
the terms $V[R_{3/2},R_{-3/2}]$ of the two orderings in Eq.~\eqref{Solution3w} cancel. The $\delta\Delta_{2\omega}$ part of
$\Xi_{RR}$ then becomes $({\cal A}_+/\omega)[X_{RR}(\eps+\omega/2)-X_{RR}(\eps-\omega/2)]$, and the same holds for the
Keldysh component (Appendix~\ref{app:third}): the $\delta\Delta_{2\omega}$ part of $\gc_3$ is the first finite difference of
the $2\omega$ response to the pairing field, in agreement with Appendix~\ref{AppendixA:gauge}. With elastic scattering
the cancellation is incomplete for two reasons. The points are displaced from $\eps+s\omega$ into the complex plane, and
the potentials at the two centre energies contain different vertex terms.

\subsection{Details of the calculation.} In this part of Appendix we provide the derivation of the results highlighted above. We use the notation of Appendix~\ref{AppendixC} for the
four energies $\eps+s\omega$ with $s=\pm\frac12,\pm\frac32$: $\gh^a_s\equiv\gh^a_{\bn,\eps+s\omega}$, $\eta^a_s$,
$\tveps^a_s$, $t_s\equiv t_{\eps+s\omega}$ and $H^a_s=\eta^a_s\gh^a_s=\tveps^a_s\ttau3+\hat\Delta_\bn$.

\paragraph{Sources.} The rotation \eqref{KeldyshRotation}, with $\check L^{-1}_{\eps+3\omega/2}$ on the left and
$\check L_{\eps-3\omega/2}$ on the right, brings Eqs.~(\ref{Eq3w},\ref{Norm3w}) to the form
\beg\label{SectorEq3}
H^a_{3/2}\,\Xi_{ab}-\Xi_{ab}\,H^b_{-3/2}=\hat S_{ab},\qquad \gh^a_{3/2}\,\Xi_{ab}+\Xi_{ab}\,\gh^b_{-3/2}=-\hat Z_{ab},
\en
whose solution is Eq.~\eqref{ClosedForm2w} with the points $a_{3/2}$ and $b_{-3/2}$. The first-order functions
$\gc_1(\bn,\eps\pm\omega)$ enter through their blocks $\gh^a_1(\eps\pm\omega)$ and $\delta\gh^K_1(\eps\pm\omega)$,
Eq.~\eqref{FirstOrderDD}, and the second-order functions $\gc_2(\bn,\eps\pm\omega/2)$ through the blocks
$X^\pm_{ab}\equiv X_{ab}(\eps\pm\omega/2)$ of Eq.~\eqref{Solution2w}. The matrix $\delta\check\sigma_2-\delta\check\Delta_\bn$
is rotated as in Appendix~\ref{AppendixC}. Its diagonal blocks are $\hat U_{RR}$ and $\hat U_{AA}$ and its off-diagonal block
is $\hat U_{RA}$, Eq.~\eqref{Potential2w}. Then
\beg\label{Sources3w}
\begin{aligned}
\hat S_{RR}&={\cal A}_+\big(X^+_{RR}\ttau3-\ttau3X^-_{RR}\big)+\hat U^+_{RR}\,\gh^R_1(\eps-\omega)-\gh^R_1(\eps+\omega)\,\hat U^-_{RR},\\
\hat Z_{RR}&=\gh^R_1(\eps+\omega)\,X^-_{RR}+X^+_{RR}\,\gh^R_1(\eps-\omega),\\
\hat S_{RA}&={\cal A}_+\big[(t_{-1/2}-t_{-3/2})X^+_{RR}\ttau3+X^+_{RA}\ttau3-\ttau3X^-_{RA}-(t_{3/2}-t_{1/2})\ttau3X^-_{AA}\big]\\
&+\hat U^+_{RR}\,\delta\gh^K_1(\eps-\omega)+\hat U^+_{RA}\,\gh^A_1(\eps-\omega)-\gh^R_1(\eps+\omega)\,\hat U^-_{RA}-\delta\gh^K_1(\eps+\omega)\,\hat U^-_{AA},\\
\hat Z_{RA}&=\gh^R_1(\eps+\omega)\,X^-_{RA}+\delta\gh^K_1(\eps+\omega)\,X^-_{AA}\\
&+X^+_{RR}\,\delta\gh^K_1(\eps-\omega)+X^+_{RA}\,\gh^A_1(\eps-\omega),
\end{aligned}
\en
and $\hat S_{AA}$, $\hat Z_{AA}$ follow from $\hat S_{RR}$, $\hat Z_{RR}$ by $R\to A$. For $\delta\hat\sigma_{ab}=0$ these are
the sources of the collisionless problem with the points shifted to $\tveps^a_s$.

\paragraph{Drive part.} For any four points the divided differences of $\gh$ obey
\beg\label{DDidentities3}
\begin{aligned}
&H_1\gh[1,2,3,4]-\gh[1,2,3,4]H_4=\gh[1,2,3]\ttau3-\ttau3\gh[2,3,4],\\
&\gh_1\gh[1,2,3,4]+\gh[1,2,3,4]\gh_4=-\gh[1,2]\,\gh[2,3,4]-\gh[1,2,3]\,\gh[3,4].
\end{aligned}
\en
Both follow from the Leibniz rule for divided differences,
$(\hat F\hat G)[1,\dots,n]=\sum_{k=1}^n\hat F[1,\dots,k]\,\hat G[k,\dots,n]$, which holds for matrix-valued $\hat F$ and
$\hat G$ because the order of the factors is kept. Applied to $H\gh-\gh H=0$, with $H[1,2]=\ttau3$ and vanishing higher
divided differences of $H$, it gives the first line. Applied to $\gh\gh=1$ it gives the second line. The identities
\eqref{DDidentities} follow in the same way. With Eq.~\eqref{FirstOrderDD} and the drive parts $X^{(0)}_{ab}$ of
$X^\pm_{ab}$, the drive parts of the sources \eqref{Sources3w} are the right-hand sides of Eq.~\eqref{DDidentities3}
multiplied by ${\cal A}_+^3$ and, in the $RA$ sector, summed over the position of the branch change with the weights of
Eq.~\eqref{DriveDD3}. By the uniqueness of Eq.~\eqref{ClosedForm2w} this proves Eq.~\eqref{DriveDD3}.

\paragraph{Potential part.} For any matrix $\hat Q$ the function $V[1,2](\hat Q)$ solves $H_1V-VH_2=\hat Q\gh_2-\gh_1\hat Q$
and $\gh_1V+V\gh_2=0$ (Appendix~\ref{AppendixC}). With $H_1-H_2=(\tveps_1-\tveps_2)\ttau3$ the divided differences
\eqref{VDD} obey
\beg\label{VDDidentities}
\begin{aligned}
&H_1V[1,2;3]-V[1,2;3]H_3=-\gh[1,2]\,\hat Q-\ttau3V[2,3],&&\gh_1V[1,2;3]+V[1,2;3]\gh_3=-\gh[1,2]\,V[2,3],\\
&H_1V[1;2,3]-V[1;2,3]H_3=\hat Q\,\gh[2,3]+V[1,2]\ttau3,&&\gh_1V[1;2,3]+V[1;2,3]\gh_3=-V[1,2]\,\gh[2,3],
\end{aligned}
\en
where all functions $V$ are taken with the same $\hat Q$. The potential part of $X^-_{RR}$ is
$V[R_{1/2},R_{-3/2}](\hat U^-_{RR})$. The terms $-\gh^R_1(\eps+\omega)\hat U^-_{RR}$ and $-{\cal A}_+\ttau3X^-_{RR}$ of
$\hat S_{RR}$ and the term $\gh^R_1(\eps+\omega)X^-_{RR}$ of $\hat Z_{RR}$, taken with this potential part, are therefore
${\cal A}_+$ times the first line of Eq.~\eqref{VDDidentities} with $\hat Q=\hat U^-_{RR}$ and the points $R_{3/2}$,
$R_{1/2}$, $R_{-3/2}$. The terms $\hat U^+_{RR}\gh^R_1(\eps-\omega)$, ${\cal A}_+X^+_{RR}\ttau3$ and
$X^+_{RR}\gh^R_1(\eps-\omega)$ correspond in the same way to the second line with $\hat Q=\hat U^+_{RR}$ and the points
$R_{3/2}$, $R_{-1/2}$, $R_{-3/2}$. This proves Eq.~\eqref{Solution3w}. In the $RA$ sector the potential parts of $X^+_{RA}$
and $X^-_{RA}$ are $V[R_{3/2},A_{-1/2}](\hat U^+_{RA})$ and $V[R_{1/2},A_{-3/2}](\hat U^-_{RA})$, the blocks
$\delta\gh^K_1(\eps\pm\omega)$ carry the differences of $t_s$, and the same identities with points on the two branches give
\beg\label{Solution3wRA}
\begin{aligned}
\Xi_{RA}=\Xi^{(0)}_{RA}+{\cal A}_+\Big\{&(t_{3/2}-t_{1/2})\,V[R_{3/2},A_{1/2};A_{-3/2}]\big(\hat U^-_{AA}\big)
+V[R_{3/2},R_{1/2};A_{-3/2}]\big(\hat U^-_{RA}\big)\\
&+V[R_{3/2};A_{-1/2},A_{-3/2}]\big(\hat U^+_{RA}\big)+(t_{-1/2}-t_{-3/2})\,V[R_{3/2};R_{-1/2},A_{-3/2}]\big(\hat U^+_{RR}\big)\Big\}.
\end{aligned}
\en
For $\Gel\to0$ the $\delta\Delta_{2\omega}$ parts of Eqs.~(\ref{Solution3w},\ref{Solution3wRA}) combine into
$({\cal A}_+/\omega)[\gc_2(\bn,\eps+\omega/2)-\gc_2(\bn,\eps-\omega/2)]$, taken with the $\delta\Delta_{2\omega}$ part of
$\gc_2$, and Eq.~\eqref{DriveDD3} together with Eq.~\eqref{g3Kansatz} gives the third finite difference of
Eq.~\eqref{FiniteDifferences}. Both statements hold for all three blocks.

\paragraph{Third harmonic of the current density.} Here we provide the details on how the expression \eqref{Trace3w} has been obtained. The subscript in \eqref{Current3w} means that only the $\ttau3$ components of $\Xi_{ab}$ are needed. By Eq.~\eqref{WardMatrix} and
$C_{ab}/(\eta^a_1+\eta^b_2)=-f[a_1,b_2]$ (Appendix~\ref{AppendixC}), $[V[1,2](\ttau3)]_{\ttau3}=-g[1,2]$ and
$[V[1,2](\itt)]_{\ttau3}=f[1,2]$, and the divided differences \eqref{VDD} inherit these relations with $g[1,2,3]$ and
$f[1,2,3]$. Every term of the third-order solution therefore contributes to the current through a divided difference of
the scalar functions $g$ or $f$. We write $\hat U_{ab}=-u^{(3)}_{ab}\ttau3-u^{(2)}_{ab}\itt$, that is
\beg\label{uComponents}
u^{(3)}_{ab}=\frac{i\Gel}{2}X^{(3)}_{ab},\qquad u^{(2)}_{ab}=w_{ab}\Ycal_\bn\delta\Delta_{2\omega}+\frac{i\Gel}{2}X^{(2)}_{ab},
\en
and introduce for an amplitude $v_{ab}(\eps)$ and an equilibrium function $Q^a_\eps$ the sum over the two orderings
\beg\label{PathSum}
\begin{aligned}
{\cal S}_{RR}[v,Q]&=v^-_{RR}\,Q[R_{3/2},R_{1/2},R_{-3/2}]\\
&+v^+_{RR}\,Q[R_{3/2},R_{-1/2},R_{-3/2}],
\end{aligned}
\en
with $v^\pm_{ab}\equiv v_{ab}(\eps\pm\omega/2)$. ${\cal S}_{AA}$ follows by $R\to A$, and ${\cal S}_{RA}$ is given in
Eq.~\eqref{PathSumRA}.

\paragraph{Components and current.} The matrices $\Xi_{ab}$ have no $\ttau0$ and $\ttau1$ components. For
$\hat Q=-u^{(3)}\ttau3-u^{(2)}\itt$ the relations stated before Eq.~\eqref{uComponents} and the symmetry of the divided
differences give $[V[1,2;3](\hat Q)]_{\ttau3}=[V[1;2,3](\hat Q)]_{\ttau3}=u^{(3)}g[1,2,3]-u^{(2)}f[1,2,3]$, which yields
Eq.~\eqref{Trace3w} with the $RA$ path sum
\beg\label{PathSumRA}
\begin{aligned}
{\cal S}_{RA}[v,Q]&=(t_{3/2}-t_{1/2})\,v^-_{AA}\,Q[R_{3/2},A_{1/2},A_{-3/2}]+v^-_{RA}\,Q[R_{3/2},R_{1/2},A_{-3/2}]\\
&+v^+_{RA}\,Q[R_{3/2},A_{-1/2},A_{-3/2}]+(t_{-1/2}-t_{-3/2})\,v^+_{RR}\,Q[R_{3/2},R_{-1/2},A_{-3/2}].
\end{aligned}
\en
The advanced divided differences follow from the retarded ones, $g[A_{s_1},\dots,A_{s_n}]=-(g[R_{s_1},\dots,R_{s_n}])^*$
and the same for $f$, since $g^A=-(g^R)^*$, $f^A=-(f^R)^*$ and $\tveps^A=(\tveps^R)^*$.
 The averages \eqref{PolAverages3} follow from
$(\bn\cdot\bE)\bn=\frac12\bE+\frac12\big(E_x\cos2\phi_\bn+E_y\sin2\phi_\bn,\,E_x\sin2\phi_\bn-E_y\cos2\phi_\bn\big)$ and from
the corresponding decomposition of $(\bn\cdot\bE)^3\bn$ into the harmonics $0$, $\pm2$ and $\pm4$ of $\phi_\bn$, since for a
function $h$ of $\Ycal_\bn^2=1+\cos4\phi_\bn$ only $\avg{h}$, $\avg{h\cos4\phi_\bn}$ and, with one factor $\Ycal_\bn$,
$\avg{h\Ycal_\bn\cos2\phi_\bn}=\avg{\Ycal_\bn^2h}/\sqrt2$ survive. At large $|\eps|$ the $RA$ terms decay exponentially. In the $RR$ and $AA$ terms $c_{RR}$ and $c_{AA}$ tend to
$\pm\,{\rm sgn}(\eps)$, and the integrand of $\Lambda_{\rm H}$ decays as $|\eps|^{-3}$, those of $A$ and $B$ as
$|\eps|^{-5}$. All energy integrals in Eqs.~(\ref{LambdaH}), (\ref{jqp3w}) and (\ref{AB3w}) converge.

\end{appendix}
\end{widetext}

\bibliography{dwave3rd}
\end{document}